\documentclass[prl,aps,preprintnumbers,twocolumn, acknowledgement]{revtex4}

\usepackage{amsmath,amssymb,bm}
\usepackage{graphicx}
\usepackage{color}

\def\eqa{\begin{eqnarray}}
\def\eea{\end{eqnarray}}
\newcommand{\eq}{\begin{equation}}
\newcommand{\ee}{\end{equation}}

\newcommand{\p}{\partial}

\newcommand{\ket}[1]{\ensuremath{\left|#1\rangle \right.}}

\newcommand\inner[2]{\ensuremath{\langle #1  | #2  \rangle }}
\newcommand\dirac[3]{\ensuremath{\langle #1 | #2 | #3 \rangle }}

\begin{document}
\title{Disorder Induced Anomalous Hall Effect in Type-I Weyl Metals:\\
Connection  between the Kubo-Streda Formula in the Spin and Chiral basis}
\author{Jia-Xing Zhang$^{1}$}
\author{Zhi-Yuan Wang$^{1}$}
\author{Wei Chen$^{1,2}$} \email{chenweiphy@nju.edu.cn}
\affiliation{$^{1}$National Laboratory of Solid State Microstructures and School of Physics, Nanjing University, Nanjing, China}
\affiliation{$^2$Collaborative Innovation Center of Advanced Microstructures, Nanjing University, Nanjing, China}

\begin{abstract}
We study the anomalous Hall effect (AHE) in tilted Weyl metals with weak Gaussian disorder under the Kubo-Streda  formalism in this work. 
To separate the three different contributions, namely the intrinsic, side jump and skew scattering contributions,
it is usually considered necessary to go to the eigenstate (chiral) basis of the Kubo-Streda formula. However, it is more straightforward to compute the total Hall current in the spin basis. For this reason, we develop a  systematic and transparent scheme to separate the three different contributions  in the spin basis for relativistic systems by building a one-to-one correspondence between the Feynman diagrams of the different mechanisms in the chiral basis and the products of the symmetric and antisymmetric parts of the polarization operator in the spin basis. 
We obtained the three contributions of the AHE in tilted Weyl metals by this scheme and found that the side jump contribution exceeds both the intrinsic and skew scattering contributions for the low-energy effective Hamiltonian.  
We compared the anomalous Hall current obtained from our scheme with the results from the semiclassical Boltzmann equation approach under the relaxation time approximation and found that the results from the two approaches agree with each other in the leading order of the tilting velocity.  
\end{abstract}

\maketitle  

\section{I. Introduction}\label{Sec:I}
The anomalous Hall effect (AHE) has attracted great interest since the starting work by Karplus and Luttinger~\cite{Luttinger1954}. 
Although it is first discussed  in ferromagnetic metals, the AHE has been discovered in various systems in the following decades, 
such as  magnetic topological insulators~\cite{Qi2008, Yu2010, Chang2013, Chang2015}, and Moire materials~\cite{Xie2020, Eom2000, Nomura2006, Sondhi1993, Yang1994, Young2012}, Dirac and Weyl metals~\cite{Wan2011, Burkov2014, Pesin2017}. All these systems include two ingredients: (i) some sort of spin- or pseudospin-orbit interaction and (ii) time reversal symmetry (TRS) breaking. 
The spin-orbit interaction results in a transverse motion of the electrons perpendicular to the electric field applied, and 
the time reversal symmetry breaking is necessary to avoid the cancellation of the  transverse flux in opposite directions~\cite{Sinitsyn2008, Yang2011}. 
The AHE  in such systems can be divided to the intrinsic contribution, which is the AHE obtained in the clean limit and is related to the topology of the electronic band structure~\cite{Niu1999, Niu2002, Onoda2002},  
and the extrinsic contribution which is due to impurity scatterings~\cite{Sinitsyn2008, Sinitsyn2007, Inoue2006, Onoda2006,  Ado2016}. 
The extrinsic Hall current can be further divided to the side jump and skew scattering contribution, according to whether the impurity scattering is symmetric or asymmetric.
  In a ferromagnetic semiconductor, the AHE is dominated by the intrinsic contribution
 and determined by the Chern number of the filled bands~\cite{Onoda2006}. In ferromagnetic metals, however, 
 the extrinsic contribution due to impurity scatterings can be significant and even cancel out~\cite{Inoue2006} or exceed the intrinsic contribution depending on the impurity scattering rate and strength~\cite{Onoda2006}.

In this work we study the AHE in a type-I Weyl metal with breaking time reversal symmetry. For simplicity, we consider the case with the minimum number of two Weyl nodes~\cite{Vishwanath2018}.
The low energy effective Hamiltonian of a typical type-I Weyl metal with tilting has the simplest form of spin-orbit interaction ${\cal H}_\chi= v \chi {\boldsymbol \sigma} \cdot {\mathbf k}+\mathbf{u_\chi\cdot k}$
 with a linear dispersion in each Weyl node.
 The anomalous Hall effect in a untilted Weyl metal, i.e., $\mathbf{u}=0$, with dilute impurities was studied in Ref.~\cite{Burkov2014}. 
 Different from two-dimensional (2D) ferromagnetic metals, the anomalous Hall effect due to impurity scatterings in the untilted Weyl metal vanishes only if  the Fermi energy is  within the linear dispersion regime of the system. This is because the low-energy Hamiltonian of the untilted Weyl metals ${\cal H}_\chi= v \chi {\boldsymbol \sigma} \cdot {\mathbf k}$
 has an emergent TRS at each single Weyl node. For this reason, the AHE in such a system then  comes completely from the  topological Chern-Simons term and is proportional to the distance between the two Weyl nodes~\cite{Burkov2014, Burkov2015, Qi2013}.

The main focus of this work is then to study the  impurity scattering induced AHE in tilted Weyl metals with dilute Gaussian disorder, which is nonvanishing since  the tilting breaks the TRS in a single Weyl node~\cite{Pesin2017}.  We  obtain the intrinsic and extrinsic contribution from the Kubo-Streda formula in the spin basis~\cite{Streda1982}. However, it is usually considered hard to separate the  side jump and the skew scattering contribution in the spin basis, since a single Feynman diagram in the spin basis contains both contributions~\cite{Sinitsyn2007}. To separate the two types of extrinsic contributions, physicists usually turn to the semiclassical Boltzmann equation (SBE) approach~\cite{Sinitsyn2006, Sinitsyn2007, Sinitsyn2008, Loss2003, Mott1929, Smit1955} since it's physically more transparent. The key ingredient resulting in the anomalous Hall current is the band mixing by the current vertex and/or impurity scatterings in the system~\cite{Luttinger1955, Luttinger1958, Sinitsyn2008}. From the SBE approach, one can separate the symmetric and anti-symmetric impurity scatterings  which  result in the side-jump and skew scattering contributions respectively. Based on the SBE approach, Sinitsyn et al. further figured out the rigorous Feynman diagrams  in the band eigenstate basis (also called the chiral basis) in the Kubo-Streda formalism corresponding to the side-jump and skew scattering contributions~\cite{Sinitsyn2007}.

Yet since the Hamiltonian of the anomalous Hall system is usually written in the spin basis due to the spin-orbit interaction, it is  cumbersome to compute the diagrams corresponding to the side jump and skew scattering contributions one by one in the chiral basis. Instead, it is  straightforward  to compute the total AHE in the spin basis.  An efficient and transparent method to separate the side jump and skew scattering contributions in the spin basis is then valuable in practice.

In this work we show that  by separating the different parts of the polarization  matrix ${\cal I}$ in the spin basis, namely the  symmetric part ${\cal I}^s$, the intrinsic antisymmetric part ${\cal I}^a_{int}$ in the clean limit and the antisymmetric part ${\cal I}^a_{im}$ due to impurity scatterings, and using these parts as building blocks, one can build up a rigorous one-to-one correspondence between the Feynman diagrams  of the side jump or skew scattering contribution in the chiral basis and the product of  matrices ${\cal I}^s, {\cal I}^a_{int}, {\cal I}^a_{im}$ in the spin basis for Weyl and Dirac systems, as shown in Fig.\ref{fig:Table}. From this correspondence, one can easily separate the intrinsic, side jump and skew scattering contributions in the spin basis. By this scheme, we separate the three contributions to the AHE in  tilted Weyl metals and found that the side jump contribution exceeds both the intrinsic and skew scattering contribution in tilted Weyl metals.

For the isotropic 2D massive Dirac systems, Sinitysn et al. shows that the AHE due to the three mechanisms obtained from the Kubo-Streda formula matches completely with the results obtained from the SBE approach~\cite{Sinitsyn2007}.  However,  for anisotropic systems, there is concern that the commonly used SBE approach under the relaxation time approximation (RTA) in Refs.~\cite{Sinitsyn2007,Loss2003} may not be reliable since 
the solution of the SBE in these works assumes that the relaxation time defined in the solution is independent of the direction of the incident electrons~\cite{Sinova2009, Sinova2009-2}. This is true for isotropic systems but not  for anisotropic systems. 
It is then interesting to compare the AHE  obtained from the SBE approach and the result obtained from the quantum Kubo-Streda formula for tilted Weyl metals, whose Fermi surface is anisotropic. We found that the RTA is still valid in the leading order of the tilting velocity $u/v$ [see Eq.(\ref{eq:tilted_Hamiltonian}) for definition] for tilted Weyl metals. As a result, the AHE obtained from the two approaches agree well with each other in the leading order of the tilting velocity for all the three mechanisms and the deviation only comes from higher orders of tilting velocity.

The structure of this paper is as follows. In Sec.II, we compute the AHE of the tilted Weyl metals using the quantum Kubo-Streda formula in the spin basis, and separate the intrinsic and extrinsic contributions. We also separate the different parts ${\cal I}^s, {\cal I}^a_{int}, {\cal I}^a_{im}$ of the polarization operator ${\cal I}$ in the spin basis. 
In Sec. III.A, we build up a one-to-one correspondence between the Feynman diagrams in the chiral basis and the matrices defined above in the spin basis for Weyl and Dirac systems. In Sec.III.B, we use this scheme to separate the side jump and skew scattering contributions in the tilted Weyl metals. In Sec.III.C, we compare the three contributions to the AHE  we obtained from the Kubo-Streda formula with the results obtained from the commonly-used SBE approach. In Sec.IV, we have a comparison of the three different approaches in studying the AHE, followed by a brief discussion of the third and fourth order crossed diagrams. We have a brief summary of this paper in Sec. V.

\section{II. Anomalous Hall current from the Kubo-streda formula in the spin basis}\label{sec:II}
We start with the effective low energy Hamiltonian of a type-I Weyl metal with breaking TRS
\begin{equation}\label{eq:tilted_Hamiltonian}
H=\sum_\chi (\chi v \boldsymbol \sigma\cdot \mathbf{k}+\mathbf {u_\chi \cdot k}),
\end{equation}
where $\chi=\pm 1$ is the chirality of the two Weyl nodes, ${\boldsymbol \sigma}$ are the Pauli matrices and $\mathbf{u}_\chi$ is a tilting velocity with $u_\chi<v$, i.e., we only consider the type-I Weyl metals. The Hamiltonian $H_\chi$ for each single valley results in two tilting  linear bands $\epsilon_{\pm}= \pm v k + \mathbf {u_\chi \cdot k}$.  Here we assume the tilting $\mathbf{u}_+=-\mathbf{u}_-=\mathbf{u}$, i.e., the tilting is  opposite for the two valleys. The tilting term then  breaks the TRS of a single Weyl node but not the global TRS of the whole system, whereas the term  $\chi v \mathbf{\boldsymbol \sigma\cdot k}$ only breaks the global TRS, but not the TRS of a single valley. We will see below that if the tilting term is the same for the two valleys with $\chi=\pm$, the AHE in the two valleys  cancels each other.

When the Weyl semi-metal or metal is coupled to an electromagnetic (EM) field $A^\alpha=(A_0, {\mathbf A})$, other than the coupling of the low energy effective Hamiltonian ]Eq.(\ref{eq:tilted_Hamiltonian})] to the EM field by the Peierls substitution of the four-momentum $p^\alpha \to p^\alpha+e A^\alpha$, an extra topological Chern-Simons term describing the chiral anomaly of the response of the Weyl semimetal ormetal to the EM field should be included in the action~\cite{Burkov2015, Qi2013}, 
\begin{equation}\label{eq:Chern_Simons}
S_\theta = -\frac{e^2}{8\pi^2} \int dt d^3 r  \partial_\mu \theta \epsilon^{\mu \nu \alpha \beta} A_\nu \partial_\alpha A_\beta,
\end{equation}
where $\epsilon^{\mu\nu\alpha\beta}$ is a {\it Levi-Civita} antisymmetric tensor, $\theta=\mathbf{b\cdot r}$ and $\mathbf{b}$ is the separation between the two Weyl nodes in the momentum space. Here we assume the energy difference between the two Weyl nodes is zero. The Chern-Simons term  results in an anomalous Hall effect with a transverse current perpendicular to both $\mathbf{b}$ and the applied electric field, 
${\mathbf j}=\frac{e^2}{2\pi^2} {\mathbf b } \times {\mathbf  E}$~\cite{Burkov2015, Qi2013}. This part turns out to be the only intrinsic contribution to the AHE of the untilted Weyl metal or semi-metal  when the doping is not very high. Moreover, the AHE in untilted Weyl semi-metal/metal is also insensitive to the impurity scatterings~\cite{Burkov2014}. For the tilted Weyl metal or semi-metal, the tilting does not affect the Chern- Simons term  since the tilting does not break the chiral symmetry. However, the  AHE for the low energy effective Hamiltonian Eq.(\ref{eq:tilted_Hamiltonian}) with tilting is no longer vanishing, and is very sensitive to the impurity scatterings. The main purpose of this work is then to study the AHE of the low energy effective Hamiltonian [Eq.(\ref{eq:tilted_Hamiltonian})] for tilted Weyl metals.

The anomalous Hall current for the low energy effective Hamiltonian can be written as two parts $j_H=j^{\mathbf I}_H+j_H^{\mathbf{II}}$ by the Kubo-Streda formalism, where $j^{\mathbf I}_H$ is a contribution from the Fermi surface and $j^{\mathbf{II}}_H$ is a contribution from the Fermi sea~\cite{Sinitsyn2007, Streda1982}.  The latter is insensitive to impurity scatterings~\cite{Niu2005, Burkov2014} and its contribution in a clean tilted  Weyl metal was studied in Ref~\cite{Pesin2017}. The contribution $j^{\mathbf I}_H$ from the Fermi surface may, however, be significantly affected by impurity scatterings. Its clean limit has been studied in Ref.~\cite{Zyuzin2017}. 
In this work we then focus on  the  AHE from the Fermi surface of the tilted Weyl metals due to  impurity scatterings, which remains poorly studied in this system.

\begin{figure}
	\includegraphics[width=8cm]{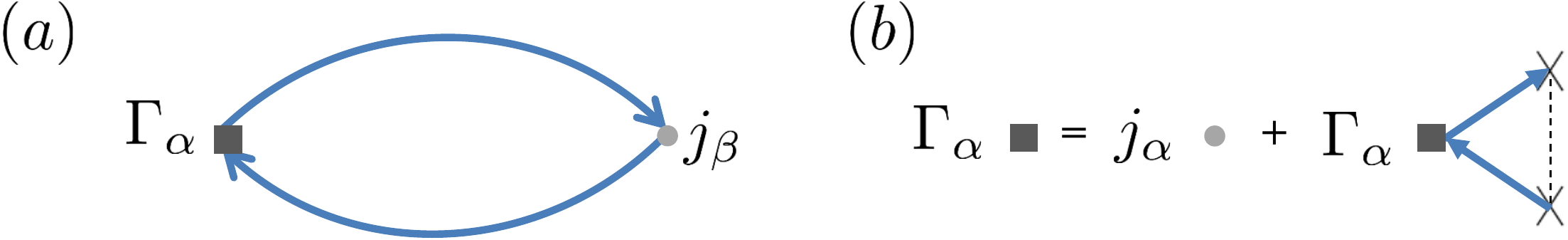}
		\caption{(a)The Feynman diagram of the response function $\Pi^{\mathbf{I}}_{\alpha\beta}$ under the non-crossing approximation (NCA) in the spin basis. The thick solid lines are Green's function in the spin basis under the Born approximation. (b)The square block is the renormalized current vertex with the NCA.}\label{fig:NCA_diagram}
\end{figure}

We consider  dilute non-magnetic impurities with  potential $V({\mathbf r})=V_0 \sum_a \delta({\mathbf r-r_a})$ and correlation $\langle V({\mathbf r})V({\mathbf r'}) \rangle=\gamma \delta({\mathbf r}-{\mathbf r'})$ where $\gamma=n_i V_0^2$ with $n_i$ the impurity density. We assume  the mean free path of the electrons is much larger than the Fermi wavelength, i.e., $k_F l \gg 1$ in this work and we mainly focus on the contribution from the non-crossing  diagram in Fig.\ref{fig:NCA_diagram} when computing the AHE. It was shown in Ref.~\cite{Ado2016, Ado2017} that the crossed diagrams become important for near impurities with distance of the order of electron Fermi wavelength. However, due to the intricacy of these diagrams, we will  study them in a separate paper.

We  assume that the impurity potential is diagonal for both the spin and valley index, so the two valleys decouple and one can compute the AHE in each valley separately and add up the contribution together at the end. In the following, we then focus on the AHE in a single Weyl node. 

The impurity averaged retarded Green's function (GF) in a single valley (e.g. with $\chi=1$ without loss of generality) under the first Born approximation is 
\begin{equation}
G^{R}(\epsilon, \mathbf{k})=(\epsilon- v \mathbf{\boldsymbol\sigma\cdot k}-\mathbf{u\cdot k} -\Sigma^R)^{-1},
\end{equation}
 where the self-energy $\Sigma^R$ due to impurity scatterings is $\Sigma^R=-\frac{i}{2\tau}[1+\sum_i\Delta_i(\mathbf{u})\sigma_i]$ with $1/\tau=\pi \gamma g(\epsilon_F)$,  
 $g(\epsilon_F)=\int \frac{d^3 k}{(2\pi)^3}\delta(\mathbf{u\cdot k} + v k -\epsilon_F)=\frac{\epsilon^2_F v}{2\pi^2(v^2-u^2)^2}$ being the density of states at the Fermi energy $\epsilon_F>0$ and
 \begin{equation}\label{eq:self_energy}
 \Delta_i(\mathbf{u})=\frac{1}{g(\epsilon_F)} \int \frac{d^3 k}{(2\pi)^3} \frac{v k_i}{\epsilon_F-\mathbf{u\cdot k} }\delta(\mathbf{u\cdot k} + v  k -\epsilon_F),
\end{equation}
where $i=x, y, z$.  
At $\mathbf{u}=0, \boldsymbol\Delta(\mathbf{u})=0$ since the integrand in Eq.(\ref{eq:self_energy})  is odd.  Whereas at finite tilting the Fermi surface is asymmetric in the momentum space and $\Delta_i=-u_i/v$.  Here $|\boldsymbol\Delta|$ describes the
difference of the scattering rates of the  spin species parallel or anti-parallel to the tilting $\mathbf{u}$. This is obvious when one chooses the direction of $\mathbf{u}$ in the $z$ direction so that only $\Delta_z=-u/v$ is non-vanishing. The difference of the impurity scattering rates of the two spin species turns out to play a key role in  the impurity scattering induced AHE as we will show below.

We consider a uniform electric field ${\mathbf E}=-\partial_t {\mathbf A}$ applied to the system. In the linear response regime $j^{\rm I}_\alpha=\Pi^{\rm I}_{\alpha\beta} A^\beta$, where $A^\alpha=(0, \mathbf{A})$, and  the leading order contribution of the response function $\Pi^{\rm I}_{\alpha\beta}$ from the Fermi surface at small frequency  under the non-crossing approximation (NCA) can be expressed by the Kubo-Streda formula as~\cite{Streda1982, Burkov2014} 
\begin{eqnarray}\label{eq:Hall_current}
&&\Pi^{\mathbf{I}}_{\alpha\beta}(\omega, {\mathbf q})=e^2 \omega\nonumber\\
 &&\int \frac{d\epsilon}{2\pi i}\frac{d{\mathbf k}}{(2\pi )^3} \partial_\epsilon n_F(\epsilon) \rm{Tr} [\hat{\Gamma}_\alpha G^R(\epsilon+\omega, \mathbf{k}+\mathbf{q})  \hat{j}_\beta G^A(\epsilon, \mathbf{k})], \nonumber\\
\end{eqnarray}
where $\hat{j}_\alpha=u_\alpha \sigma_0 + \chi v \sigma_\alpha, \alpha=0, x, y, z $ is the bare current vertex for Hamiltonian Eq.(\ref{eq:tilted_Hamiltonian}), (we define $u_0=0$), $G^{R/A}$ is the retarded (R) or advanced (A) Green's function under the first Born approximation, and $\hat{\Gamma}_\alpha$ is the renormalized current vertex due to impurity scatterings as shown in Fig.1b.  Note that the linear response function $\Pi^{\mathbf{I}}_{\alpha\beta}$ also contains the $G^RG^R$ and $G^A G^A$ terms~\cite{Sinitsyn2007}, but these terms are smaller in a factor of $1/k_F l$ than the $G^RG^A$ term in the limit $k_F l \gg1$ so we neglect it in Eq.(\ref{eq:Hall_current}).

The renormalized current vertex $\hat{\Gamma}_\alpha$ satisfies the recursion equation
\begin{equation}\label{eq:vertex_equation}
\hat{\Gamma}_\alpha(\omega, \mathbf{q})=\hat{j}_\alpha+ \int \frac{d^3k}{(2\pi)^3}\gamma  G^A(0, \mathbf{k}) \hat{\Gamma}_\alpha G^R(\omega, \mathbf{k+q}).
\end{equation}

The above equation can be solved by expressing the  current vertex in the Pauli matrix basis as $\hat{j}_\alpha={\cal J}_{\alpha \beta} \sigma_\beta, \alpha, \beta=0, x, y, z$, and the renormalized current vertex as $\hat{\Gamma}_\alpha=\Gamma_{\alpha \beta} \sigma_\beta$, where the summation over the repeated index  is implied as usual.
For Hamiltonians with the coefficients ${\cal J}_{\alpha\beta}$ and $\Gamma_{\alpha\beta}$  independent of the momentum, such as the Weyl and Dirac systems, 
 the coefficients of the renormalized current vertex can be solved, with the relationship $\rm{Tr} [ \sigma_\alpha \sigma_\beta]=2\delta_{\alpha\beta}$,  as $\Gamma_{\alpha \beta} ={\cal J}_{\alpha \gamma}D_{\gamma \beta} $, and ${\cal D}=(1-\gamma{\cal I})^{-1}$  is the $4\times 4$ diffusion matrix with the polarization operator ${\cal I}$ defined as
\begin{equation}\label{eq:I_matrix_0}
{\cal I}_{\alpha\beta}=
\frac{1}{2}\int \frac{d{\mathbf k}}{(2\pi )^3}\rm{Tr} [\sigma_\alpha G^R(\epsilon+\omega, \mathbf{k}+\mathbf{q})  \sigma_\beta G^A(\epsilon, \mathbf{k})], 
\end{equation}
and $\alpha, \beta=0, x, y, z$. 
Note that for the Hamiltonians with non-relativistic kinetic energy term ${\mathbf k}^2/2m$, such as 2D Rashba ferromagnets~\cite{Ado2016}, ${\cal J}_{\alpha\beta}$ is dependent on the momentum and the solution of renormalized current vertex is not so simple. We focus on the case with ${\cal J}_{\alpha\beta}$ independent of the momentum in this work, such as the relativistic systems, and our study in this work is applicable only  for such systems.

The $4\times4$ response function matrix $\Pi^{\rm I}(\omega, \mathbf{q})$  with vertex correction can then be expressed as the following matrix product
\begin{equation}\label{eq:total_response}
\tilde{\Pi}^{\rm I}(\omega, \mathbf{q})={\cal J}{\cal D}{\cal I} {\cal J}^T,
\end{equation}
where  $\tilde{\Pi}^{\rm I} \equiv - \frac{\pi i}{\omega e^2}\Pi^{\rm I}$, and  ${\cal J}$ is the $4\times4$ matrix with elements ${\cal J}_{\alpha\beta}$, i.e., the coefficients  of the bare current vertex in the Pauli matrix basis as defined above. The superscript $T$ means the transposition of the corresponding matrix. The energy $\epsilon$ in $\cal I$ is bounded to the Fermi energy $\epsilon_F$ due to the factor 
$\partial_\epsilon n_F(\epsilon)$ in Eq.(\ref{eq:Hall_current}).

We are interested in the anomalous Hall conductivity in the dc limit of the diffusive regime, i.e.,  $ql \ll1, \omega \tau\ll1$. 
However, under the first Born approximation, the matrix ${\cal D}^{-1}=1-\gamma {\cal I}$  in the exact dc limit $\omega=0, \bf{q}=0$ has determinant zero and is not invertible.  For the reason, we work at small finite  frequency $\omega$ (but keep ${\bf q}$ zero for simplicity) and take the dc limit $\omega \to 0$ at the end of the calculation.

We separate the ${\cal I}$ matrix to the symmetric and anti-symmetric part as ${\cal I}(\omega, \mathbf{q} \to 0)={\cal I}^s(\omega)+{\cal I}^a(\omega)$. Here ${\cal I}^s(\omega)$ and ${\cal I}^a(\omega)$ in the diffusive limit $\omega \tau\ll 1$ can be obtained by expanding the small frequency.
In the linear order of the frequency $\omega$,  the symmetric part ${\cal I}^s(\omega)$ and the anti-symmetric part ${\cal I}^a(\omega)$ are respectively 
\begin{widetext}
\begin{equation}\label{eq:I_s}
\gamma {\cal I}^s(\omega)=\frac{1}{ c_0}\left(
\begin{array}{cccc}
  c_0+\frac{u^2}{v}c_1+i\omega \tau b_0 & c_1(\omega)u_x & c_1(\omega) u_y & c_1(\omega) u_z \\
  c_1(\omega) u_x & c_2(\omega)+c_3(\omega) u^2_x & c_3(\omega) u_x u_y & c_3(\omega) u_x u_z\\
  c_1(\omega) u_y & c_3(\omega) u_x u_y & c_2(\omega) +c_3(\omega) u^2_y & c_3(\omega) u_y u_z \\
  c_1(\omega) u_z & c_3(\omega) u_x u_z & c_3(\omega) u_y u_z & c_2(\omega) +c_3(\omega) u^2_z
\end{array}\right),\\
\end{equation}
\end{widetext}
and 
\begin{eqnarray}
\gamma {\cal I}^a(\omega) &=& \gamma {\cal I}^{a}_{int}(\omega) +\gamma {\cal I}^{a}_{im}(\omega), \\
\gamma {\cal I}_{int, \alpha\beta}^{a}(\omega) &=&\frac{1}{c_0\epsilon_F\tau} f(\omega)\epsilon^{\alpha\beta\gamma} u^\gamma,  \label{eq:anti_sym_int}\\
\gamma {\cal I}_{im,\alpha\beta}^{a}(\omega) &=&\frac{1}{c_0\epsilon_F\tau}g(\omega)\epsilon^{\alpha\beta\gamma} \Delta^\gamma.  \label{eq:anti_sym_im}
\end{eqnarray}
In the above expressions, $u^\gamma=(0, u_x, u_y, u_z)$ and $\Delta^\gamma=(0, \Delta_x, \Delta_y, \Delta_z)$. 
The matrix ${\cal I}_{int, \alpha\beta}^{a}(\omega)$ is the anti-symmetric part of ${\cal I}$ in the clean limit $\tau\to \infty$, and  ${\cal I}_{im,\alpha\beta}^{a}$ is the anti-symmetric part due to impurity scatterings.
The parameters $c_i(\omega)= c_i+i\omega\tau b_i, i=1, 2, 3$, where $c_i$ is the value of $c_i(\omega)$ in the dc limit and $b_i$ is the coefficient of the linear order  frequency term. The exact value of  $c_i$ can be obtained by directly computing the ${\cal I}$ matrix in the dc limit as in the Appendix A, whereas the coefficients $b_i$ can be obtained by expanding ${\cal I}(\omega, \mathbf q\to 0)$ to the linear order of $\omega$.  For the anomalous Hall conductivity in the dc limit, it turns out that only the  values in the dc limit of $c_i(\omega)$ matter at the end and the coefficients $b_i$ do not enter the final results of the dc anomalous Hall conductivity. The same is true for  $f(\omega)$ and $g(\omega)$. For the reason, we  only give these parameters in the dc limit in the following:
 \begin{eqnarray}
&&c_0= \frac{v^4}{(v^2-u^2)^2}, \label{eq:C_0} \\
&&b_0=\frac{1}{4 }[3+5c_0-3\frac{u^2}{v^2}c_0 +3 a(u)\frac{u^2}{v^2}], \\
&&c_1(\omega\to 0)\equiv c_1= \frac{1}{4v}[a(u)+\frac{u^2-3v^2}{v^2}c_0],\label{eq:C_1}\\
& &c_2(\omega\to 0)\equiv c_2 
=\frac{1}{8}[3+\frac{2u^2}{v^2-u^2}-\frac{v^2-u^2}{v^2}a(u)],   \label{eq:C_2}\\
 && c_3(\omega\to 0)\equiv c_3=\frac{1}{8  u^2}[3a(u)-1+4\frac{u^2}{v^2}c_0-\frac{u^2}{v^2}a(u)],\label{eq:C_3} \nonumber\\
&& f(\omega\to 0)\equiv f(u)= \frac{1}{2v}[\frac{v^2}{v^2-u^2}-a(u)], \label{eq:f_u}\\
&&  g(\omega\to 0)\equiv g(u)=-\frac{1}{4}[1+a(u)+\frac{u^2}{v^2}a(u)],\label{eq:g_u}\\
&& a(u)\equiv \frac{v^2}{u^2}(\frac{v}{2u}{\rm ln}\frac{v+u}{v-u}-1)\approx\frac{1}{3}+\frac{u^2}{5v^2}+{\cal O}(\frac{u^4}{v^4}). \nonumber\\
\end{eqnarray}

In the above calculation, when the chirality $\chi$ changes sign but $\mathbf{u}$ does not, both $f(u)$ and $\mathbf{\Delta}$ changes sign. When $\mathbf{u}$ changes sign but $\chi$ does not, $f(u)$ does not change sign, but $\mathbf{\Delta}$ does. The parameter $g(u)$ does not change sign when $\chi$ or ${\mathbf u}$ changes sign. So the sign of the anti-symmetric part ${\cal I}^a$ is determined by the product of the sign of $\chi$ and $\mathbf{u}_\chi$ in each valley. In this work, $\chi$ is opposite in the two valleys since the total chiral charge of the two valleys has to be zero~\cite{Vishwanath2018}. The tilting $\mathbf{u}_\chi$ then must have opposite signs in the two valleys to get non-vanishing total AHE as we will see below.

From the $\cal{I}$ matrix, we can obtain the diffusion matrix $\cal{D}$ and the renormalized current vertex $\hat{\Gamma}$ and finally get the Hall current through Eq.(\ref{eq:total_response}). The main contributions to the AHE in a tilted Weyl metal are presented as follows.

{\it AHE without vertex correction.} In this case, ${\cal D}=1$, the response function 
\begin{equation}\label{eq:Pi_no_vertex}
\Pi^{\rm I}(\omega, \mathbf{q})=-\frac{\omega}{\pi i}e^2  {\cal J}{\cal I}(\omega, \mathbf{q}){\cal J}^T.
\end{equation}
For tilted Weyl metals, ${\cal J}_{\alpha\beta}=v \delta_{\alpha\beta}+u_\alpha \delta_{\beta 0}$. It is easy to check that in the dc limit  the tilting part $u_\alpha \delta_{\beta, 0}$ in the bare current vertex of the tilted Weyl metals has no contribution to the AHE in the linear response and 
$ \Pi^{\rm I}_{\alpha\beta}(\omega, \mathbf{q})=-\frac{\omega}{\pi i}e^2  v^2 {\cal I}_{\alpha\beta}(\omega, \mathbf{q})$ without vertex correction
 for the tilted Weyl metals. The main effect of the tilting is then just to produce an anisotropy in the Fermi surface.

For isotropic systems, the spatial block of the symmetric part ${\cal I}^s$, i.e., ${\cal I}^s_{ij}, i, j=x, y, z,$ is diagonal and the diagonal element corresponds to the longitudinal conductivity that is proportional to $\tau$~\cite{Loss2003}. For anisotropic systems, ${\cal I}^s_{ij}, i, j=x, y, z,$ may contain off-diagonal elements, as is the case for the tilted Weyl metals in Eq.(\ref{eq:I_s}). This off-diagonal part corresponds to a normal Hall current due to anisotropy~\cite{Loss2003}.
 The anti-symmetric part ${\cal I}^a$ of the ${\cal I}$ matrix comes from TRS breaking and corresponds to an anomalous Hall current.
 This part is  smaller than the symmetric part ${\cal I}^s$ by a factor of $1/\epsilon_F\tau$, consistent with the fact that the anomalous Hall conductivity is usually smaller than the longitudinal conductivity by  $1/\epsilon_F\tau$~\cite{Sinitsyn2007}.

 The  anti-symmetric part ${\cal I}^a$ of ${\cal I}(\omega, \mathbf{q})$  in the limit of $\omega\to 0, \mathbf{q}\to 0$ results in a dc anomalous Hall current as 
\begin{equation}\label{eq:AHE_no_vertex}
\mathbf{j}_H=\frac{e^2 \epsilon_F }{2\pi^2v}[f(u) (\mathbf{u\times E})+g(u)(\mathbf{\Delta \times E})],
\end{equation}
where $f(u)$ and $g(u)$ are given in Eqs.(\ref{eq:f_u}) and (\ref{eq:g_u}), and $\Delta_i=-\frac{u_i}{v}, i=1,2,3$.

Equation (\ref{eq:AHE_no_vertex}) above contains two terms. The first term, which we denote as $\mathbf{j}^{int}_H$, corresponds to the clean and dc limit of  the anti-symmetric tensor ${\cal I}$, i.e., ${\cal I}^a_{int}(\omega\to 0)$ in Eq.(\ref{eq:anti_sym_int}), and together with the contribution $\mathbf{j}^{\rm II}_H$ from the Fermi sea constitute the  intrinsic contribution of the low energy effective Hamiltonian.  Note that the contribution corresponding to ${\cal I}^a_{int}(\omega\to 0)$ vanishes in the Weyl metals without tilting, 
as can be seen from Eq.(\ref{eq:I_matrix_0}) where the integration over $\mathbf{k}$ and $-\mathbf{k}$ for the anti-symmetric  tensor ${\cal I}^a_{\alpha\beta}$ cancels out at $\omega, \mathbf{q} \to 0$ for this case.

The second term in Eq.(\ref{eq:AHE_no_vertex}) corresponds to ${\cal I}^a_{im}(\omega\to 0)$ in Eq.(\ref{eq:anti_sym_im}).
This contribution is due to an unbalanced impurity scattering of the spin species parallel and anti-parallel to the tilting direction $\mathbf{u}$. It  is proportional to the difference of the impurity scattering rate  of the two spin species, but is independent of the total  impurity scattering rate $1/\tau$.
This contribution involves a second order impurity scattering process $|V_0|^2$ and constitutes part of  the side jump contribution as we will show in more details in the next section.

{\it Vertex correction.} We next consider the AHE in tilted Weyl metals with vertex correction. Since the anti-symmetric part ${\cal I}^a$ is smaller than the symmetric part ${\cal I}^s$ by a factor of $1/\epsilon_F\tau$, we can expand the diffusion matrix $\cal{D}$ treating $\gamma{\cal I}^a$ as a small parameter,
\begin{equation}
{\cal D}=(1-\gamma {\cal I}^s-\gamma{\cal I}^a)^{-1}=\sum_{n=0}^\infty ({\cal D}_0 \gamma {\cal I}^a)^n {\cal D}_0,
\end{equation}
 where $\gamma {\cal I}^a\sim 1/\epsilon_F\tau$, and ${\cal D}_0\equiv (1-\gamma {\cal I}^s)^{-1}$ is a constant symmetric matrix $\sim \tau^0$ and satisfying ${\cal D}_0=1+{\cal D}_0\gamma {\cal I}^s$. The matrix ${\cal D}_0$ gives the leading order vertex correction of the current operator.
 
 For the vertex correction coming from the anti-symmetric part $\gamma {\cal I}^a$ in the ${\cal D}$ matrix, one only needs to keep the lowest order of this contribution since $\gamma {\cal I}^a$ is smaller than $\gamma {\cal I}^s$ by $1/\epsilon_F\tau$.
For tilted Weyl metals, the response function  $\tilde{\Pi}^{\rm I}={\cal J}{\cal D} {\cal I}{\cal J}^T=v^2{\cal D} {\cal I}$.
Upon expansion, we get the anti-symmetric part for tilted Weyl metals as
\begin{eqnarray}\label{eq:vertex_correction}
\tilde{\Pi}^{\rm I}_a/v^2&=&[{\cal D}_0+{\cal D}^2_0(\gamma {\cal I}^a)^2 {\cal D}_0+...]{\cal I}^a, \nonumber\\
&&+[{\cal D}_0(\gamma {\cal I}^a){\cal D}_0+{\cal D}^3_0(\gamma {\cal I}^a)^3{\cal D}_0+...]{\cal I}^s,\nonumber\\ 
&=& {\cal D}_0{\cal I}^a+{\cal D}_0(\gamma {\cal I}^a){\cal D}_0{\cal I}^s+...
\end{eqnarray}

The leading order contribution  contains two terms, i.e.,  ${\cal D}_0{\cal I}^a$ and ${\cal D}_0(\gamma {\cal I}^a){\cal D}_0 {\cal I}^s$. Both of the   terms are independent of $\tau$. The addition of the two terms is equal to ${\cal D}_0 {\cal I}^a {\cal D}_0$ for tilted Weyl metals, and ${\cal J}{\cal D}_0 {\cal I}^a {\cal D}_0{\cal J}^T$ for general anomalous Hall systems with ${\cal J}$ independent of the momentum.

To get the matrix ${\cal D}_0$, one needs to invert the matrix $1-\gamma {\cal I}^s$. However, in the dc limit $\omega=0, {\mathbf q}=0$, the matrix $1-\gamma {\cal I}^s$  has determinant zero due to charge conservation~\cite{Burkov2014} and is not invertible. This can be verified by checking the determinant $\rm Det[1-\gamma {\cal I}^s(\omega\to 0)]=-\frac{u^2}{v}\frac{c_1}{c_0^4}(c_0-c_2)^2(c_0-c_2-c_3 u^2 + v c_1)$, for which the factor $c_0-c_2-c_3 u^2 + v c_1$ equals to zero exactly.
However, at finite frequency, the matrix $1-\gamma {\cal I}^s$ becomes invertible. From Eq.(\ref{eq:I_s}), we get
\begin{widetext}
\begin{eqnarray}
  &\mathcal{D}_0(\omega)=\frac{c_0}{[c_0-c_2(\omega)]\mu(\omega)} \left(
\begin{array}{cccc}
  c_0-c_2(\omega)-c_3(\omega) u^2 & c_1(\omega)u_x & c_1(\omega)u_y & c_1(\omega)u_z\\
 c_1(\omega)u_x & \mu(\omega)+\lambda(\omega) u_x^2 & \lambda(\omega)u_x u_y &\lambda(\omega)u_x u_z\\
  c_1(\omega)u_y & \lambda(\omega)u_x u_y &\mu(\omega)+\lambda(\omega) u_y^2 & \lambda(\omega)u_y u_z \\
  c_1(\omega)u_z & \lambda(\omega)u_x u_z & \lambda(\omega)u_y u_z &\mu(\omega)+\lambda(\omega) u_z^2 \\
\end{array}\right),
\end{eqnarray} 
where 
\begin{eqnarray}
&&\mu(\omega)= [(-\frac{u^2}{v}c_1-i\omega \tau b_0)(c_0-c_2(\omega)-c_3(\omega) u^2)-c_1^2(\omega)u^2] /[c_0-c_2(\omega)],\\
    && \lambda(\omega)= [c_1^2(\omega)-c_3(\omega)(\frac{u^2}{v}c_1+i\omega \tau b_0)]/[c_0-c_2(\omega)].
\end{eqnarray}    
\end{widetext}

One can  check that in the leading order of $\omega\tau$, the parameter $\mu(\omega)\propto i\omega \tau$, resulting in a diffusion pole at $\omega\to 0$ in the diffusion matrix ${\cal D}_0$. 
 It is interesting to get the leading order renormalized current vertices from  ${\cal D}_0$ for the tilted Weyl metals, which are  
\begin{widetext}
\begin{eqnarray}
&&\Gamma_0=\Gamma_{0 \beta}\sigma_{\beta}=\frac{v c_0}{[c_0-c_2(\omega)]\mu(\omega)}[(c_0-c_2(\omega)-c_3(\omega) u^2)\sigma_0 +c_1(\omega)\boldsymbol u \cdot \boldsymbol\sigma],  \label{eq:Gamma_0}\\
   &&\Gamma_i=\Gamma_{i \beta}\sigma_{\beta}=\frac{ c_0}{c_0-c_2(\omega)}[\frac{c_0-c_2(\omega)-c_3(\omega) u^2 +v c_1(\omega)}{\mu(\omega)}u_i\sigma_0 
   +v \sigma_i+ \frac{c_1(\omega)+v \lambda(\omega)}{\mu(\omega)}u_i(\boldsymbol u \cdot \boldsymbol\sigma)], i=1,2,3. \label{eq:Gamma_i}
\end{eqnarray}
\end{widetext}
It is clear from Eq.(\ref{eq:Gamma_0}) that the renormalized charge component $\Gamma_0$ has a diffusion pole at $\omega\to 0$ due to the charge conservation. 
However, since both $c_0-c_2(\omega)-c_3(\omega) u^2 +v c_1(\omega)$ and $c_1(\omega)+v \lambda(\omega)$ in Eq.(\ref{eq:Gamma_i}) vanish at $\omega\to 0$ and are proportional to $i\omega\tau$ in the leading order of the frequency, the renormalized current vertices $\Gamma_i, i=1, 2, 3$ have no diffusion pole. This is because the current operators $\hat{j}_i, i=1, 2,3$ do not commute with the Hamiltonian and are not conserved.

From the ${\cal D}_0$ and ${\cal I}^a$ matrix, we get the leading order contribution of $\tilde{\Pi}^{\rm I}_a(\omega)$ for tilted Weyl metals as
\begin{eqnarray}
\tilde{\Pi}^{\rm I}_a(\omega)&=&v^2{\cal D}_0(\omega) {\cal I}^a(\omega) {\cal D}_0(\omega) \nonumber\\
&=&[\frac{c_0}{c_0-c_2(\omega)}]^2v^2 \mathcal{I}^a(\omega).
\end{eqnarray}

Taking the limit $\omega\to 0$, we get the total dc anomalous Hall current  for tilted Weyl metals in a single valley as  
\begin{equation}\label{eq:total_Hall_current}
{\mathbf j}^{\rm I}_H =\frac{e^2 \epsilon_F}{2\pi^2v}F(u)[f(u) (\mathbf{u\times E})
 + g(u)(\mathbf{\Delta \times E})]   ,
\end{equation}
where $F(u)=\frac{c^2_0}{(c_0-c_2)^2}$ with $c_0, c_2$ given in Eqs.(\ref{eq:C_0}) and (\ref{eq:C_2}), and $f(u)$ and $g(u)$  given in Eqs.(\ref{eq:f_u}) and (\ref{eq:g_u}). At ${\mathbf u}=0, F(u)=\frac{c^2_0}{(c_0-c_2)^2}=9/4$, and at finite $\mathbf{u}$, the value of $F(u)$ is plotted in Fig.2.

\begin{figure}
	\includegraphics[width=7.5cm]{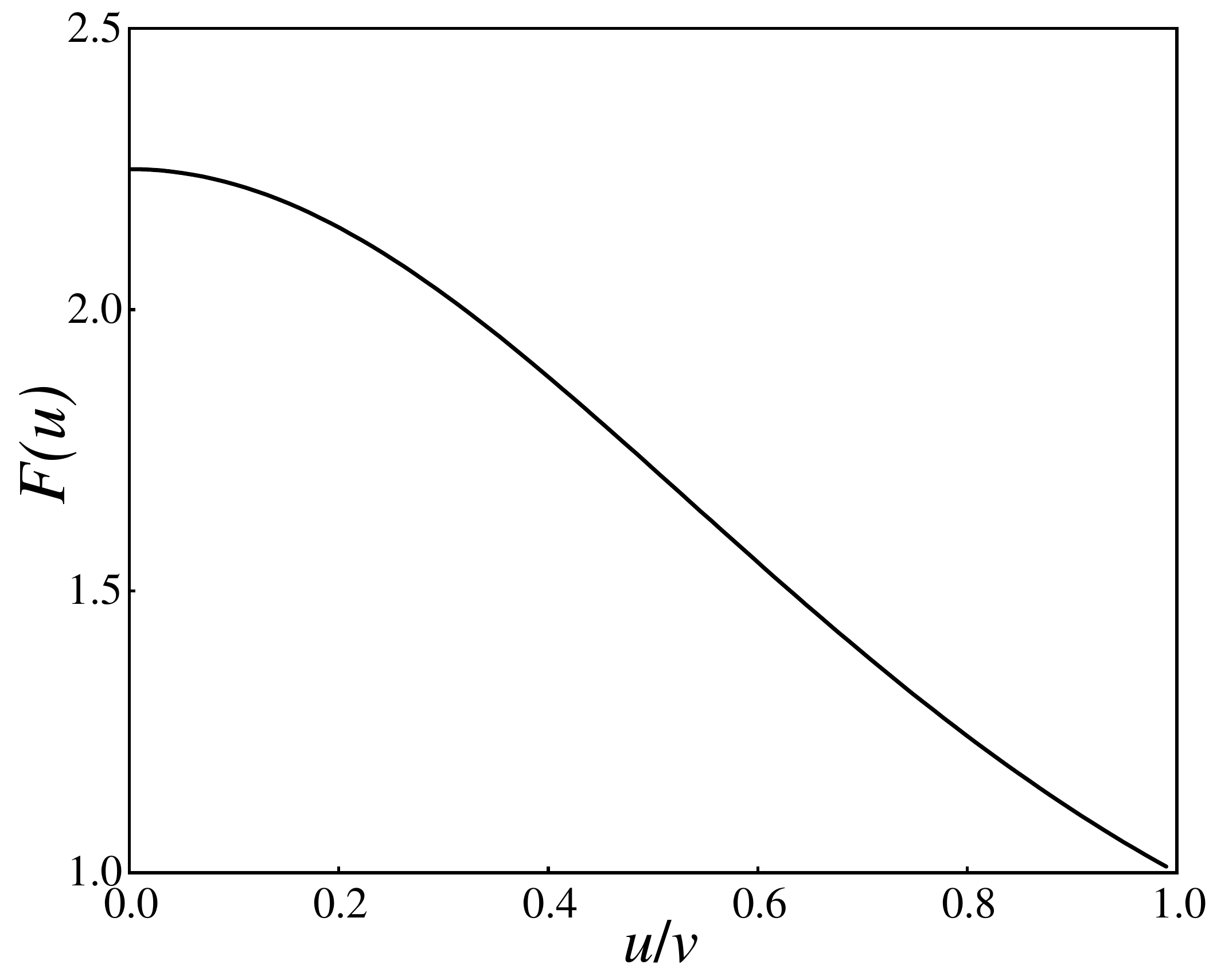}
		\caption{The factor $F(u)$ for the vertex correction as a function of $u/v$. }\label{fig:F_u}
\end{figure}

From Eq.(\ref{eq:vertex_correction}), the next leading order vertex correction to the Hall current is $\sim {\cal D}^3_0(\gamma {\cal I}^a)^2 {\cal I}^a+{\cal D}^4_0(\gamma {\cal I}^a)^3 {\cal I}^s\sim (1/\epsilon_F\tau)^2$,  which is much smaller than the leading order correction in the case $\epsilon_F\tau\gg1$. For this reason, we only need to keep the leading order contribution of the AHE which is of the order of $\tau^0$.

The total anomalous Hall current in Eq.(\ref{eq:total_Hall_current}) includes both the intrinsic and  extrinsic contribution. The intrinsic one corresponds to the first term of Eq.(\ref{eq:AHE_no_vertex}). The total contribution subtracting the intrinsic one is then the extrinsic contribution due to impurity scatterings, which includes the side-jump and skew scattering contribution.  In the next section we show how to separate these two contributions from the total Hall current in Eq.(\ref{eq:total_Hall_current}).

\section{III. Separation of the intrinsic, side-jump and skew scattering contributions in the spin basis}\label{sec:III}

\subsection{A. General Formalism}

The Hall current in Eq.(\ref{eq:total_Hall_current}) obtained from the Kubo-Streda formula in the spin basis is rigorous but not physically transparent. It took physicists a long time to understand the microscopic mechanism of  both the intrinsic and  extrinsic contribution of the AHE. With the efforts of the authors in Refs.~\cite{Sinitsyn2006, Smit1955, Berger1970, Luttinger1955, Luttinger1958, Sinitsyn2007, Sinitsyn2008} and others, a semi-classical Boltzmann equation approach was built to explain the anomalous Hall current. The key ingredient is the band mixing by the current vertex and/or the impurity potential which induces a transverse current perpendicular to the electric field. Based on the semi-classical explanation of the three different mechanisms of the anomalous Hall currents, i.e., the intrinsic, 
side-jump and  skew scattering,
Sinitsyn et al. figured out the Feynman diagrams corresponding to the three different mechanisms in the band eigenstate  basis (shown in Fig.\ref{fig:Table})~\cite{Sinitsyn2007}.  The intrinsic contribution is due to the topological structure of the electron band and exists even without impurity scatterings. The side-jump contribution is due to the  transverse displacement of the electrons by impurity scatterings~\cite{Berger1970}. This scattering is symmetric. Whereas the skew scattering is due to asymmetric impurity scatterings.

Though it is physically more transparent to separate the different contributions in the eigenstate or chiral basis, as done in most previous works~\cite{Sinitsyn2007,Yang2011, Sinova2010},  the Hamiltonians of the anomalous Hall systems are usually given in the spin basis due to the spin-orbit interaction. For this reason, it is  more convenient to calculate the total  anomalous Hall current in the spin basis as we did in the last section. However, a transparent method to separate the different extrinsic contributions in the spin basis is lacking and considered to be a difficult task  since a single diagram in the spin basis  contains contributions from different mechanisms~\cite{Sinitsyn2007, Yang2011}.

In this section, we show an efficient and transparent scheme to separate the different contributions of the AHE in the spin basis for Dirac and Weyl systems. By separating the symmetric and anti-symmetric part of the $\cal{I}$ matrix and $\cal{D}$ matrix as we did in the last section and using these matrices as building blocks, we build a one-to-one correspondence between the Feynman diagrams of the different contributions given in the chiral basis in Ref~\cite{Sinitsyn2007} and the matrix products of  ${\cal I}^{a}_{int}, {\cal I}^{a}_{im}$ and ${\cal D}_0$ we obtained  in the spin basis. The result is shown in Fig.\ref{fig:Table}. For simplicity, we build up this correspondence for tilted Weyl metals at first and it is easy to generalize the result to other Weyl and Dirac systems.

For tilted Weyl metals, the tilting part $e u_\alpha\sigma_0$ in the bare current vertex has no contribution to the linear response. The ${\cal I}$ matrix remains the same by replacing $\sigma_\alpha$ by $\hat{j}_\alpha/v$.
The  integrand
  of the ${\cal I}$ matrix in Eq. (\ref{eq:I_matrix_0}) can then be replaced by 
$I_{\alpha\beta}={\rm Tr}[\hat{j}_\alpha G^{R}\hat{j}_{\beta} G^A]$ for tilted Weyl metals. To get better physical understanding, we use the latter notation to expand $I_{\alpha\beta}$ in this section. The result can be easily generalized to other systems with different current vertices independent of the momentum.

The integrand $I_{\alpha\beta}$ is the same in  the spin and chiral bases. 
We expand it in the chiral basis as 
\begin{eqnarray}\label{eq:integrand}
&&I_{\alpha\beta}= \nonumber\\
&&\langle+| \hat{j}_\alpha \sum_{s1} |s_1\rangle \langle s_1| G^R \sum_{s2} |s_2\rangle \langle s_2| \hat{j}_{\beta} \sum_{s3} |s_3\rangle \langle s_3|G^A |+\rangle \nonumber\\
&&+
\langle -| \hat{j}_\alpha \sum_{s4} |s_4\rangle \langle s_4| G^R \sum_{s5} |s_5\rangle \langle s_5| \hat{j}_{\beta} \sum_{s6} |s_6\rangle \langle s_6|G^A |-\rangle,\nonumber\\
\end{eqnarray}
where $s_{1,...,6}=\pm$ represent the upper and lower eigenbands respectively. The Green's function under the Born approximation can be written as 
\begin{equation}\label{eq:Born_approximation}
G^{R/A}=G_0^{R/A}+G_0^{R/A}\Sigma G^{R/A},
\end{equation}
where $G_0^{R/A}$ is the bare GF without impurity scatterings, and $\Sigma$ is the self-energy due to impurity scatterings given in the last section. In the eigenstate basis, $G_0^{R/A}$ is diagonal, but $\Sigma$ contains both diagonal and off-diagonal elements, i.e., $\Sigma$ may cause band mixing. For the reason, $G^{R/A}$ in the Born approximation also contains off-diagonal elements in the band eigenstate basis. However, its off-diagonal elements are small in $1/\tau$. For this reason, 
we only need to keep at most one band off-diagonal element of $G^R$ or $G^A$ in each term in Eq.(\ref{eq:integrand}). The leading order terms are then 
\begin{eqnarray}\label{eq:I_integrand}
I_{\alpha\beta}&=&\dirac{+}{\hat{j}_\alpha}{+}\dirac{+}{G^R}{+}\dirac{+}{\hat{j}_\beta}{+}\dirac{+}{G^A}{+} \nonumber\\
\indent \ \   &+&\dirac{-}{\hat{j}_\alpha}{-}\dirac{-}{G^R}{-}\dirac{-}{\hat{j}_\beta}{-}\dirac{-}{G^A}{-} \nonumber\\
 \indent  \ \    &+&\dirac{+}{\hat{j}_\alpha}{-}\dirac{-}{G^R}{-}\dirac{-}{\hat{j}_\beta}{+}\dirac{+}{G^A}{+} \nonumber\\
 \indent  \ \    &+&\dirac{-}{\hat{j}_\alpha}{+}\dirac{+}{G^R}{+}\dirac{+}{\hat{j}_\beta}{-}\dirac{-}{G^A}{-} \nonumber\\
 \indent  \ \    &+&\dirac{+}{\hat{j}_\alpha}{-}\dirac{-}{G^R}{+}\dirac{+}{\hat{j}_\beta}{+}\dirac{+}{G^A}{+} \nonumber\\
  \indent  \ \  &+&\dirac{+}{\hat{j}_\alpha}{+}\dirac{+}{G^R}{-}\dirac{-}{\hat{j}_\beta}{+}\dirac{+}{G^A}{+} \nonumber\\
 \indent  \ \    &+&\dirac{+}{\hat{j}_\alpha}{+}\dirac{+}{G^R}{+}\dirac{+}{\hat{j}_\beta}{-}\dirac{-}{G^A}{+} \nonumber\\
 \indent  \ \    &+&\dirac{-}{\hat{j}_\alpha}{+}\dirac{+}{G^R}{+}\dirac{+}{\hat{j}_\beta}{+}\dirac{+}{G^A}{-}\nonumber\\
   \indent \ \    &+&\dirac{-}{\hat{j}_\alpha}{+}\dirac{+}{G^R}{-}\dirac{-}{\hat{j}_\beta}{-}\dirac{-}{G^A}{-} \nonumber\\
\indent \ \    &+&\dirac{-}{\hat{j}_\alpha}{-}\dirac{-}{G^R}{+}\dirac{+}{\hat{j}_\beta}{-}\dirac{-}{G^A}{-} \nonumber\\
 \indent  \ \    &+&\dirac{+}{\hat{j}_\alpha}{-}\dirac{-}{G^R}{-}\dirac{-}{\hat{j}_\beta}{-}\dirac{-}{G^A}{+} \nonumber\\
 \indent  \ \    &+&\dirac{-}{\hat{j}_\alpha}{-}\dirac{-}{G^R}{-}\dirac{-}{\hat{j}_\beta}{+}\dirac{+}{G^A}{-}.
 \end{eqnarray}
 
\begin{figure}
	\includegraphics[width=8cm]{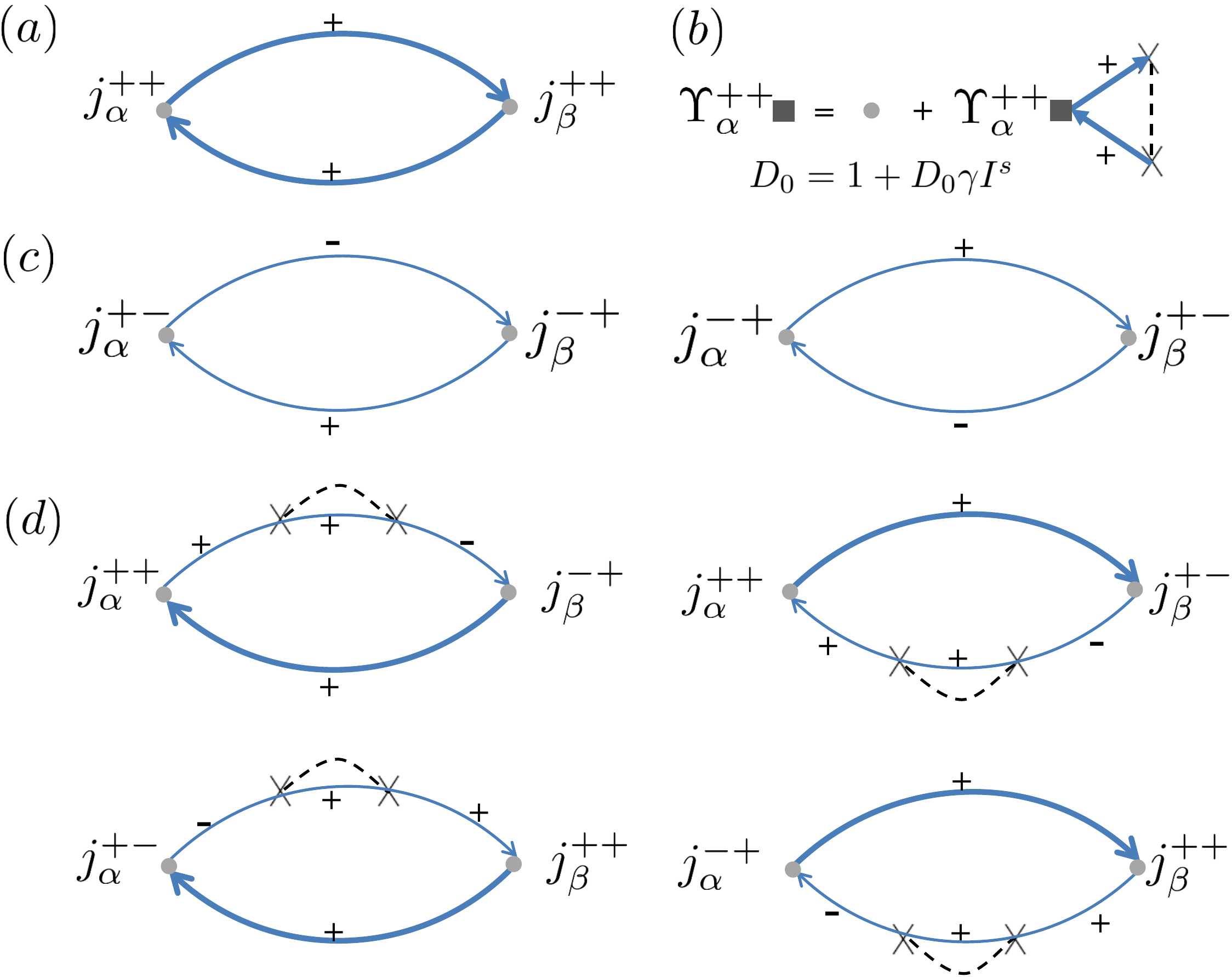}
		\caption{Correspondence of the symmetric and anti-symmetric part of the ${\cal I}$ matrix in the spin basis to the Feynman diagrams in the band eigenstate basis for tilted Weyl metals.  The thin solid lines represent the bare GF $G_0$ in the eigenstate basis and the thick solid lines represent the GF under the first Born approximation in the eigenstate basis. The dashed lines represent impurity scatterings. The diagram in (a) corresponds to the symmetric part ${\cal I}^s$ of the ${\cal I}$ matrix. The solid square in (b) represents the renormalized current vertex in the chiral basis. It corresponds to the renormalization by the symmetric part ${\cal D}_0$ of the diffusion matrix ${\cal D}$ in the spin basis. The matrix ${\cal D}_0$ satisfies the recursion relationship ${\cal D}_0=1+{\cal D}_0 \gamma {\cal I}^s$ and causes an intraband vertex correction of the current vertex. The two diagrams in (c) represent the intrinsic antisymmetric part ${\cal I}^{a}_{int}$ of the ${\cal I}$ matrix. The four diagrams in (d) represent the antisymmetric part of the ${\cal I}$ matrix due to impurity scatterings ${\cal I}^{a}_{im}$. }\label{fig:I_diagram}
\end{figure}

\begin{figure}
	\includegraphics[width=8cm]{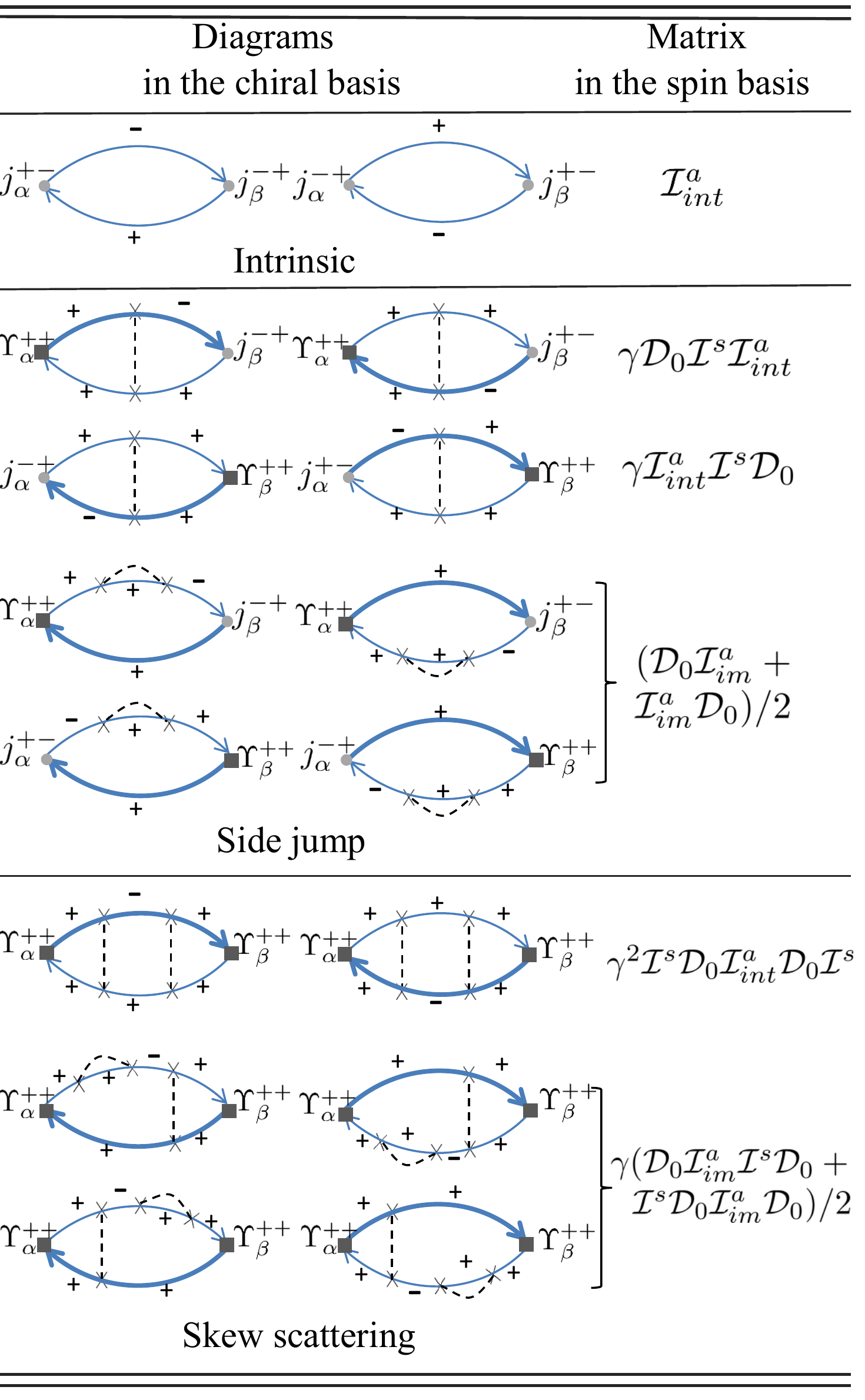}
		\caption{The correspondence of the Feynman diagrams  in the eigenstate (chiral) basis for the intrinsic, side jump and skew scattering contributions to the matrix product of ${\cal I}^s, {\cal D}_0, {\cal I}^a_{int}, {\cal I}^a_{im}$ obtained in the spin basis for tilted Weyl metals. }\label{fig:Table}
\end{figure}

 We   separate these terms into three groups:
 
  (i) The first two terms correspond to intraband processes, and both current vertices are band diagonal without mixing the two bands. As shown in the Appendix B, these two terms are the dominant symmetric part of the $I$ matrix.
We denote them as $I^s$. Among the two terms in $I^s$, the first term with intraband processes in the upper band dominates the contribution to the integration, since the lower band is far below the Fermi surface and the scattering processes by current vertex are forbidden. For the reason,  the symmetric part ${\cal I}^s$ of the ${\cal I}$ matrix corresponds to the diagram  in Fig.~\ref{fig:I_diagram}(a) in the chiral basis.

(ii) The next two terms involve only band diagonal elements of $G^R$ and $G^A$, but both current vertices mix the two bands. For these two terms, we further separate the part in the clean limit and the remaining part with disorder scattering by Eq.(\ref{eq:Born_approximation}). It is easy to show that the part in the clean limit dominant (see the Appendix B). This part corresponds to ${\cal I}^{a}_{int}$ we obtained in the spin basis since it is the only nonvanishing antisymmetric term in the clean limit in the band eigenstate basis. It then gives the intrinsic contribution to the AHE from the Fermi surface and  corresponds to the diagram in Fig.  \ref{fig:I_diagram}(c) in the band basis.

(iii)The remaining eight terms in $I_{\alpha\beta}$ are also antisymmetric in the leading order and are nonvanishing only with impurity scatterings. The four terms with $\langle -|G^{R/A} |-\rangle$ in the lower band have an extra smallness compared to the other four terms and so are negligible.  For the remaining four terms, one can expand the band off-diagonal Green's function elements $\langle \pm|G^{R/A} |\mp\rangle$ by Eq.(\ref{eq:Born_approximation}). Since $\langle \pm|G_0^{R/A} |\mp\rangle =0$,  $\langle \pm|G^{R/A} |\mp\rangle=\langle \pm|G_0^{R/A}\Sigma G^{R/A} |\mp\rangle \approx \langle \pm|G_0^{R/A}\Sigma G_0^{R/A} |\mp\rangle$, where the last replacement of $G^{R/A}$ by $G_0^{R/A}$ in this element does not affect the integration of $I^{\alpha\beta}$ in the leading order. The leading order contribution of this part then corresponds to the four diagrams in Fig.\ref{fig:I_diagram}d and the matrix ${\cal I}^{a}_{im}$ we obtained in the spin basis.

Note that for all the diagrams in Figs.\ref{fig:I_diagram}(c) and Fig.\ref{fig:I_diagram}(d), replacing the thin line representing $G^{R/A}_0$ by the thick line representing $G^{R/A}$ does not change the leading order contribution of the AHE. For clarity of physics, we only keep the minimum number of thick lines in the leading order in this paper. The same applies to the diagrams in Fig.\ref{fig:Table}.

The symmetric part ${\cal I}^s$ does not contribute to the Hall current directly since it involves only intraband scattering. However, it produces a band-diagonal  vertex correction of the current operator through the symmetric part of the diffusion matrix, i.e.,  ${\cal D}_0$ that  we defined in the last section. 
The renormalized band-diagonal current vertex is labeled as $\Upsilon^{++}$ in the band eigenstate basis in Ref~\cite{Sinitsyn2007}, and satisfies the recursion relationship shown in Fig. \ref{fig:I_diagram}b.

The antisymmetric part of diffusion operator ${\cal D}$ is smaller than the symmetric part  by a factor $1/\epsilon_F\tau$ so the vertex correction due to this part does not need a resummation to infinite order. This part generates a transverse component of the current vertex and one only needs to keep its lowest order as we did in the last section.

The Feynman diagrams of the intrinsic, side-jump and skew scattering contribution to the AHE in the leading order of $1/\epsilon_F\tau$  in the band eigenstate basis are given in Ref~\cite{Sinitsyn2007}, also shown in Fig.\ref{fig:Table}.  With the  correspondence in Fig. \ref{fig:I_diagram} between the diagrams in the band eigenstate  basis and the matrices in the spin basis, we can easily write down the matrices for each diagram of the three different mechanisms for tilted Weyl metals. The results are also presented in Fig.\ref{fig:Table}.  From this table, we can conveniently separate the intrinsic, side  jump and skew scattering contribution in the spin basis for tilted Weyl metals.

The intrinsic contribution corresponds to the ${\cal I}$ matrix in the clean limit, i.e., ${\cal I}^{a}_{ int}$. The side jump contribution involves only second order impurity scatterings, i.e., only one disorder line. There are eight diagrams in the eigenstate (chiral) basis for this type, as shown in Fig.\ref{fig:Table}~\cite{Footnote0}. Their total contribution to the AHE is then 
\begin{eqnarray}\label{eq:side_jump_1}
\tilde{\Pi}^{sj} &=&\gamma {\cal D}_0 {\cal I}^s {\cal I}^a_{int} + {\cal I}^a_{int} \gamma {\cal D}_0 {\cal I}^s+\frac{{\cal D}_0 {\cal I}^a_{im}+ {\cal I}^a_{im}{\cal D}_0}{2} \nonumber\\
&=& \frac{{\cal D}_0 {\cal I}^a+{\cal I}^a {\cal D}_0}{2} -{\cal I}^a_{int} \nonumber\\
&+&\frac{({\cal D}_0-1) {\cal I}^a_{int}+{\cal I}^a_{int}({\cal D}_0-1)}{2}. 
\end{eqnarray}
where  $\tilde{\Pi}^{sj} \equiv -\frac{\pi i}{\omega e^2 v^2}\Pi^{sj}$ is the rescaled response function as in the last section.  One may wonder how to equate the matrix $\gamma {\cal D}_0 {\cal I}_s {\cal I}^a_{int}$, i.e., $( {\cal D}_0-1) {\cal I}^a_{int}$ with the corresponding diagrams on the left in Fig.\ref{fig:Table} since
$ {\cal D}_0-1$ only connects to upper band legs whereas  the vertices of ${\cal I}^a_{int}$ mix the upper and lower bands. This is easy to understand by noting that $( {\cal D}_0-1) {\cal I}^a_{int}$ is the leading order contribution of $({\cal D}-1) {\cal I}^a_{int}$ and the disorder line of $({\cal D}-1)$ includes all possible  inter-band and intra-band scatterings by disorder. Among all the diagrams corresponding to $({\cal D}-1) {\cal I}^a_{int}$, the leading order contribution only corresponds to the diagrams on the left of Fig.\ref{fig:Table}.

The second line of Eq.(\ref{eq:side_jump_1}) equals  the leading order contribution with only one  vertex correction ${\cal D}_0$ in Eq.(\ref{eq:vertex_correction}), subtracting the intrinsic contribution ${\cal I}^a_{int}$. Other than this term, the side jump contribution contains another term  $\frac{({\cal D}_0-1) {\cal I}^a_{int}+{\cal I}^a_{int}({\cal D}_0-1)}{2}$, which comes from the second term in Eq.(\ref{eq:vertex_correction}). The remaining part of the  second term in Eq.(\ref{eq:vertex_correction}) is the contribution from skew scattering as we show below.

The skew scattering contribution corresponds to the six diagrams in Fig.\ref{fig:Table}. Each diagram involves two disorder lines, i.e., the fourth order impurity scattering processes. We can read the total contribution to the AHE from the skew scattering diagrams as 
\begin{eqnarray}\label{eq:skew_scattering_1}
\tilde{\Pi}^{sk}&=& \gamma {\cal D}_0 {\cal I}^s {\cal I}^a_{int} \gamma {\cal D}_0 {\cal I}^s \nonumber\\
&& + \frac{{\cal D}_0 {\cal I}^a_{im} \gamma {\cal D}_0 {\cal I}^s + \gamma {\cal D}_0 {\cal I}^s {\cal I}^a_{im} {\cal D}_0}{2}, \nonumber\\
&=& \frac{({\cal D}_0-1) {\cal I}^a {\cal D}_0+ {\cal D}_0 {\cal I}^a ({\cal D}_0-1)}{2} \nonumber\\
&&-\frac{({\cal D}_0-1) {\cal I}^a_{int}+{\cal I}^a_{int}({\cal D}_0-1)}{2}. \label{eq:skew_scattering}
\end{eqnarray}
The first term of Eq.(\ref{eq:skew_scattering_1}) corresponds to the antisymmetric part of the term ${\cal D}_0 {\cal I}^a {\cal D}_0  \gamma{\cal I}_s$ in Eq.(\ref{eq:vertex_correction}), which contains two vertex corrections ${\cal D}_0$. However, the antisymmetric part of  ${\cal D}_0 {\cal I}^a {\cal D}_0  \gamma{\cal I}_s$ is not fully the skew scattering contribution. It contains part of the side jump contribution as subtracted in the last line.

It is easy to check that the addition of the intrinsic contribution, the side jump contribution in Eq.(\ref{eq:side_jump_1}) and the skew scattering contribution in Eq.(\ref{eq:skew_scattering_1}) is equal to the total Hall response  $\tilde{\Pi}^I_a={\cal D}_0{\cal I}^a {\cal D}_0$ for tilted Weyl metals obtained in the spin basis in Eq.(\ref{eq:vertex_correction}).

The above results can be easily generalized to other anomalous Hall systems without nonrelativistic kinetic energy term, such as a 2D massive Dirac model~\cite{Sinitsyn2007}. For a general anomalous Hall system with ${\cal J}$ independent of the momentum, one only needs to multiply the matrix ${\cal J}$ and ${\cal J}^T$ on the two ends of the response function matrix,
as well as the side jump contribution in Eq.(\ref{eq:side_jump_1}) and skew scattering contribution in Eq.(\ref{eq:skew_scattering_1}). The total contribution to the Hall response is $\tilde{\Pi}^I_a={\cal J}{\cal D}_0 {\cal I}^a {\cal D}_0{\cal J}^T$.

\subsection{B. The vertex correction factor ${\cal D}_0$}
We have a brief discussion of the  vertex correction of the current operator in Fig.\ref{fig:I_diagram}b in this section. The renormalized current vertex due to this vertex correction is related to the bare current vertex by $\Gamma_{\alpha\beta}\sigma_\beta={\cal J}_{\alpha \gamma}{\cal D}_{0, \gamma\beta}\sigma_\beta$ in the spin basis  as we show in the last section. Here ${\cal D}_0$ is the symmetric part of the diffusion matrix ${\cal D}$ in the dc limit.

 As discussed in the last section, this vertex correction corresponds to only the impurity scatterings within the upper band in the ladder diagram Fig.\ref{fig:I_diagram}(b) in the chiral  basis.
 In the chiral basis, the renormalized band-diagonal current vertex $\Upsilon^{++}_\alpha$ is obtained from the recursion equation as~\cite{Sinitsyn2007}
\begin{equation}\label{eq:classical_eq}
\Upsilon^{++}_\alpha({\mathbf k})=j^{++}_\alpha({\mathbf k})+ \int \frac{d^3 {\mathbf k'}}{(2\pi)^3}G^{A+}|V^{++}_{{\mathbf k'} \mathbf{k}}|^2 G^{R+} \Upsilon^{++}_\alpha({\mathbf k'}),
\end{equation}
where $j^{++}_\alpha(\mathbf k)=\langle \mathbf k, +| \hat{j}_\alpha| \mathbf k, +\rangle$, $G^{R/A+}=\langle {\mathbf k'},+|G^{R/A}|{\mathbf k'},+\rangle$, $V_{{\mathbf k'}{\mathbf k}}^{++}=\langle {\mathbf k'},+|V|{\mathbf k},+\rangle$.
For the case with isotropic Fermi surface and isotropic scattering potential, the scatterings  only depend on the angle between ${\mathbf k'}$ and ${\mathbf k}$, and the above equation can be solved exactly by the ansatz 
\begin{equation}\label{eq:classical_vertex}
\Upsilon^{++}_\alpha({\mathbf k})=\tilde{\alpha} j^{++}_\alpha({\mathbf k}),
\end{equation}
where $\tilde{\alpha}$ is a constant depending on the angle between $\mathbf{k}$ and $\mathbf{k}'$ but not the direction of $\mathbf{k}$. For untilted Weyl metals, as well as many other isotropic systems, such as graphene and the 2D massive Dirac model,  the band-diagonal current vertex in the chiral basis is $j^{++}_\alpha\sim k_\alpha$ for $\alpha=x, y, z$. For these cases,  $\tilde{\alpha}$ can be  solved as $\tilde{\alpha}=\tau^{tr}/\tau^+$, where $\tau^{tr}$ and $\tau^+$ are the transport and ordinary scattering times of the upper band respectively and can be expressed as
\begin{equation}\label{eq:scattering_rate_tr}
1/\tau^{tr}=2\pi  \int \frac{d^3 {\mathbf k'}}{(2\pi)^3}|V^{++}_{{\mathbf k'} \mathbf{k}}|^2 (1-\cos({\mathbf k'}\cdot{\mathbf k}))\delta(\epsilon_{\mathbf{k}}-\epsilon^+_{\mathbf k'}), 
\end{equation}
\begin{equation}\label{eq:scattering_rate_+}
1/\tau^+=2\pi  \int \frac{d^3 {\mathbf k'}}{(2\pi)^3}|V^{++}_{{\mathbf k'} \mathbf{k}}|^2 \delta(\epsilon_{\mathbf k}-\epsilon^+_{\mathbf k'}).
\end{equation}

For these isotropic systems, the vertex correction obtained from the above procedure in the chiral basis should be equivalent to that obtained from ${\cal D}_0$ in the spin basis. Indeed, for the spin-orbit interacting systems with isotropic Fermi surface and isotropic impurity scatterings, the matrix ${\cal D}_0$ is completely diagonal if the Hamiltonian also has time reversal symmetry, e.g., graphene~\cite{Chen2012, Ando1998, Falko2008}, a helical metal on the surface of topological insulators~\cite{Burkov2010}, or a single node of untilted Weyl metals~\cite{Burkov2014}.
And for the isotropic systems with breaking time reversal symmetry,  such as 2D massive Dirac model~\cite{Sinitsyn2007}, ${\cal D}_0$ is diagonal only for the spatial block  with $ i, j=x, y, z$.
For all these cases, the vertex correction factor ${\cal D}_0$ for the current vertex $\hat{j}_i, i=x, y$ can be replaced by  the diagonal $(i, i)$ element of ${\cal D}_{0}$, and we have checked that this element is equal to $\tilde{\alpha}$ obtained from Eqs.(\ref{eq:classical_vertex})-(\ref{eq:scattering_rate_+}). The side jump and skew scattering contribution in Eqs.(\ref{eq:side_jump_1}) and (\ref{eq:skew_scattering_1}) are then greatly simplified for such systems, and one only needs to compute the parameter $\tilde{\alpha}$ and the anti-symmetric part ${\cal I}_{int}^a$ and ${\cal I}^a_{im}$ of the polarization operator to obtain the three different anomalous Hall contributions.

However, for the systems with anisotropic Fermi surface, such as tilted Weyl metals, the impurity scatterings depend not only on the angle between ${\mathbf k'}$ and ${\mathbf k}$, but also the direction of $\mathbf{k}$. This can be seen by computing $1/\tau^{tr}$  for the tilted Weyl metals in Eq.(\ref{eq:ani_tau_tr}) in the Appendix C. It is easy to check that it depends on the direction of $\mathbf{k}$ for finite tilting. 
In this case, 
the integration equation (\ref{eq:classical_eq}) for the renormalized current vertex is intricate and does not have the simple solution as Eq.(\ref{eq:classical_vertex}). 
On the other hand, the band-diagonal vertex correction matrix ${\cal D}_0$ we obtained in the spin basis for tilted Weyl metals is no longer diagonal in the spatial block since ${\cal I}^s$ is not diagonal, indicating the complication of the vertex correction for anisotropic systems.

For the untilted Weyl metals, from our calculation in the last section, ${\cal D}_0$ is diagonal and ${\cal D}_0(i, i)=3/2$ for $i=x, y, z$, which is the same as $\tilde{\alpha}=\tau^{tr}/\tau^+$ for this system ($1/\tau^{tr}=n_i V^2_0 \epsilon^2_F/3\pi v^3, 1/\tau^{+}=n_i V^2_0 \epsilon^2_F/2\pi v^3$ for untilted Weyl metals). For the tilted Weyl metals, ${\cal D}_0$ contains both diagonal and off-diagonal elements, and both play a role in the vertex correction as we will see below.

\subsection{C. Separation of the intrinsic, side jump and skew scattering contribution in the tilted Weyl metals}
For the tilted Weyl metals, from the formalism in Sec. III.A and the matrices ${\cal D}_0, {\cal I}^a_{int}, {\cal I}^a_{im}$ we obtained in  Sec.II, we get the response functions for the three different mechanisms from a single valley as
\begin{eqnarray}
\Tilde{\Pi}^{int}(\omega)&=& \mathcal{I}^a_{int}(\omega),\\
\Tilde{\Pi}^{sj}(\omega)&=&\frac{2c_2(\omega)}{c_0-c_2(\omega)}\mathcal{I}^a_{int}(\omega)+\frac{c_0}{c_0-c_2(\omega)}\mathcal{I}^a_{im}(\omega), \\
\Tilde{\Pi}^{sk}(\omega)&=& 
   [\frac{c_2(\omega)}{c_0-c_2(\omega)}]^2\mathcal{I}^a_{int}(\omega) + \frac{c_0c_2(\omega)}{[c_0-c_2(\omega)]^2} \mathcal{I}^a_{im}(\omega).  \nonumber\\
\end{eqnarray}

Taking the limit $\omega\to 0$, we get the response functions in the dc limit as 
\begin{eqnarray}
&&\Tilde{\Pi}^{int}_{\alpha\beta}=\frac{\epsilon_F}{4\pi v^4} [\frac{v^2}{v^2-u^2}-a(u)]\epsilon^{\alpha\beta\gamma} u_{\gamma}, \label{eq:intrinsic_0}\\
    &&\Tilde{\Pi}^{sj}_{\alpha\beta}=\frac{\epsilon_F}{8\pi v^4} \frac{c_0}{(c_0-c_2)} [1+a(u)+\frac{u^2}{v^2}a(u)\nonumber\\
   &&\ \ \ \ \ \ \ \ \ \ \ \ \ \ +4\frac{c_2}{c_0}(\frac{v^2}{v^2-u^2}-a(u))]\epsilon^{\alpha\beta\gamma} u_{\gamma}, \label{eq:side_jump_0}\\
    &&\Tilde{\Pi}^{sk}_{\alpha\beta}=\frac{\epsilon_F}{8\pi v^4} \frac{c_0 c_2 }{(c_0-c_2)^2} [1+a(u)+\frac{u^2}{v^2}a(u)\nonumber\\
   &&\ \ \ \ \ \ \ \ \ \ \ \ \ \ \ +2\frac{c_2}{c_0}(\frac{v^2}{v^2-u^2}-a(u))]\epsilon^{\alpha\beta\gamma} u_{\gamma}.
   \label{eq:skew_scattering_0}    
\end{eqnarray}
From above, we see that the final results of the response functions only depend on the parameters $c_0, c_2$ and $a(u)$ obtained in the dc limit  in Sec.II.

\begin{figure}
	\includegraphics[width=8cm]{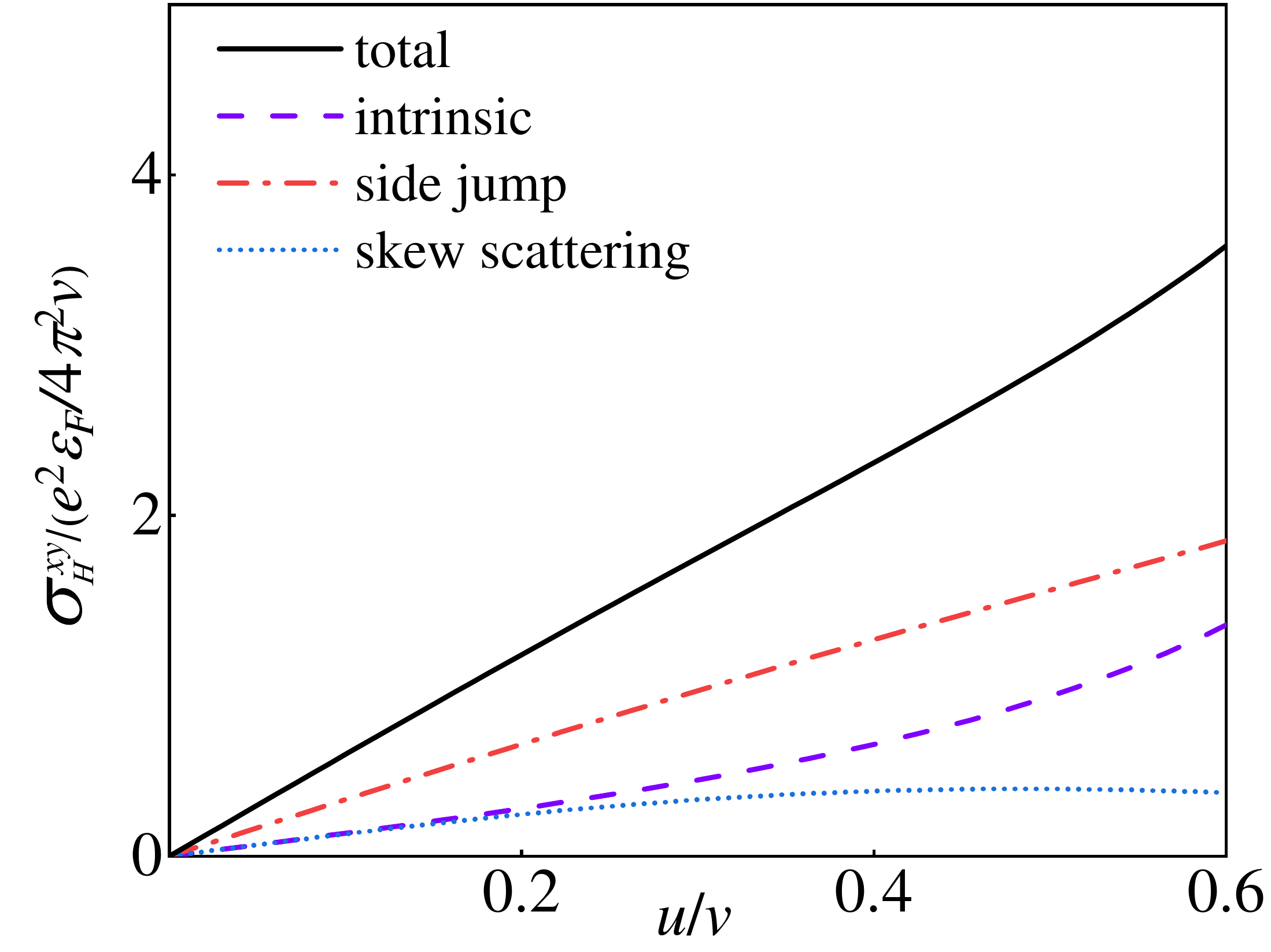}
		\caption{The Fermi surface contribution of the anomalous Hall conductivity due to the three mechanisms: intrinsic, side jump and skew scattering, obtained from the Kubo-Streda formula as a function of $u/ v$, when the electric field $\mathbf{E}$ is in the $y$ direction and $\mathbf{u}$ is in the $z$ direction. }\label{fig:Three_contributions}
\end{figure}

In the above calculation we noticed that though the tilting ${\mathbf u}$ brings correction to both the diagonal and off-diagonal elements of the ${\cal D}_0$ matrix compared to the untilted case,  the effects on the Hall currents due to the off-diagonal elements of ${\cal D}_0$ were canceled by the effects from part of the diagonal elements. The total effect of the matrix ${\cal D}_0$ on the Hall current is equivalent to the factor $c_0/(c_0-c_2)\approx 3/2-9u^2/10v^2$. However, this factor is different from $\tilde{\alpha}=\tau^{tr}/\tau^+$ for tilted Weyl metals, which is $\tilde{\alpha}\approx 3/2-3u/2v$. This also indicates that for tilted Weyl metals with anisotropic Fermi surface, the simple solution $\tilde{\alpha}=\tau^{tr}/\tau^+$ for the vertex correction is no longer accurate. 

From the response function in Eqs.(\ref{eq:intrinsic_0})-(\ref{eq:skew_scattering_0}), we can get the dc anomalous Hall conductivity $\sigma^i_{\alpha\beta}=\frac{e^2 v^2}{\pi}\tilde{\Pi}^i_{\alpha\beta}, i= {\rm int, sj, sk}$. The anomalous Hall currents from the three different mechanisms are all proportional to $\mathbf{u\times E}$ and the total anomalous Hall current from one valley is the same as obtained in Eq.(\ref{eq:total_Hall_current}). In Fig.\ref{fig:Three_contributions}, we plot the anomalous Hall conductivity due to the three different mechanisms from the Fermi surface.
 We  can see that in all regimes of $u/v$,  the side jump contribution is the largest and the extrinsic contribution is greater than the intrinsic contribution.

At small $u/v$, we can expand the results in Eqs.(\ref{eq:intrinsic_0})-(\ref{eq:skew_scattering_0}) and get the leading order  Hall currents for the two valleys as 
\begin{eqnarray}
j^{int}_H&\approx&-\frac{1}{3}\frac{e^2\epsilon_F}{\pi^2v^2}\mathbf{u\times E}, \label{eq:intrinsic}\\
j^{sj}_H &\approx&-\frac{5}{6}\frac{e^2\epsilon_F}{\pi^2v^2}\mathbf{u\times E}, \label{eq:side_jump}\\
j^{sk}_H&\approx&-\frac{1}{3}\frac{e^2\epsilon_F}{\pi^2v^2} \mathbf{u\times E}.\label{eq:skew_scattering}
\end{eqnarray}

There is another contribution to the intrinsic anomalous Hall current due to tilting  from the Fermi sea. This contribution was calculated in Ref.~\cite{Pesin2017}. In the dc limit, this Hall current is $j^{int, {\rm II}}_{H}\approx \frac{1}{6}\frac{e^2\epsilon_F}{\pi^2v^2}\mathbf{u\times E}$ for the two valleys, which is half of the intrinsic Hall current from the Fermi surface and has opposite sign. The total intrinsic Hall current in the leading order of $u/v$ due to tilting is then $j^{int, {\rm tot}}_{H}\approx-\frac{1}{6}\frac{e^2\epsilon_F}{\pi^2v^2}\mathbf{u\times E}$.

\subsection{D. Comparison with the semiclassical Boltzmann equation approach}

The AHE in the tilted Weyl metals due to  the three different mechanisms was calculated from the SBE approach in Ref.\cite{Fu2021}. However, in this work, the authors neglected the anisotropy of the Fermi surface of the system when computing the transverse coordinate shift, which results in an incorrect side jump velocity. For this reason, we redid the calculation of the SBE approach for the AHE of tilted Weyl metals in the Appendix C.

Another issue of the SBE approach is that the commonly used solution of the SBE approach under the relaxation-time approximation in Refs.~\cite{Sinitsyn2007, Loss2003} may become unreliable for anisotropic system, as pointed out in Refs.~\cite{Sinova2009, Sinova2009-2}. The reason is that the solution of the nonequilibrium distribution function $g_\mathbf{k}$ in the SBE approach assumes that the relaxation times $\tau^{tr}$ and $\tau^\bot$ do not depend on the direction of the momentum $\mathbf{k}$ (see Appendix C). This is true for isotropic systems but not the case for anisotropic systems. Strictly speaking, no scalar relaxation time can be attributed to a given $\mathbf{k}$ state for anisotropic systems. The same problem comes up for the solution of the anomalous distribution function $g^a_\mathbf{k}$ due to the coordinate shift. In Ref.\cite{Sinova2009} the authors studied the anisotropic magnetoresistance (AMR) in 2D Rashba ferromagnets with anisotropic magnetic impurities. It was shown that the exact result of the AMR in the anisotropic system is significantly different from the result obtained from the SBE under the commonly used relaxation time approximation. For this reason, it is interesting to check whether the AHE from the commonly used SBE approach also deviates significantly from the result obtained from the quantum Kubo-Streda formula for tilted Weyl metals, whose Fermi surface is anisotropic.

As shown in Appendix C, we get 
the AHE  in the leading order of $u/v$ from the SBE approach  under the relaxation time approximation after correcting the side jump velocity in Ref.\cite{Fu2021} as
\begin{eqnarray}
j^{int}_H&\approx&-\frac{1}{6}\frac{e^2\epsilon_F}{\pi^2v^2}\mathbf{u\times E}, \\
j^{sj}_H &\approx&-\frac{5e^2\epsilon_F}{6\pi^2v^2}\mathbf{u\times E}, \\
j^{sk}_H&\approx&-\frac{e^2\epsilon_F}{3\pi^2v^2} \mathbf{u\times E}.
\end{eqnarray}

The intrinsic anomalous Hall current from the SBE approach includes  contribution from both the Fermi sea and the Fermi surface~\cite{MacDonald2006}. This part is the same as the total intrinsic anomalous Hall current we obtained from the quantum Kubo-Streda formula in the last  section. Moreover, the side jump and skew scattering contributions we obtained from the SBE approach also agree with the result from the quantum Kubo-Streda formula in the leading order of $u/v$. The reason for this full match between the two approaches is because the relaxation time for tilted Weyl metals defined in Eq.(\ref{eq:ani_tau_tr})-(\ref{eq:tau_bot_tr}) is a constant independent of the momentum in the zeroth order of tilting and depends on the momentum only at higher orders of $u/v$. For this reason, the solution of the SBE  under the 
 relaxation time approximation is still valid in the leading order of $u/v$ in tilted Weyl metals, and the AHEs obtained from the SBE and Kubo-Streda formula agree with each other in the leading order of $u/v$.

\section{IV. Discussion}

For some of the anisotropic systems, it is still possible to solve the SBE exactly by introducing momentum dependent relaxation time, as shown in Ref. \cite{Sinova2009}. However, the difficulty of solving the SBE in this way is greatly enhanced because this approach involves solving two integral equations of the distribution function and there is no general solution for different models.
On the other hand, 
one encounters a similar problem for anisotropic systems when solving the vertex correction of the current operator from the recursion equation in the chiral basis, as shown in Sec. III.B. For anisotropic systems, the only convenient approach to get the rigorous anomalous Hall current for Gaussian disordered systems is then to apply the Kubo-Streda formula in the spin basis, since one can solve the recursion equation for the vertex correction in this basis exactly.  Our scheme in this work to separate the contributions from the three different mechanisms in the spin basis of the Kubo-Streda formula is then especially important for anisotropic systems.

\begin{figure}
	\includegraphics[width=8cm]{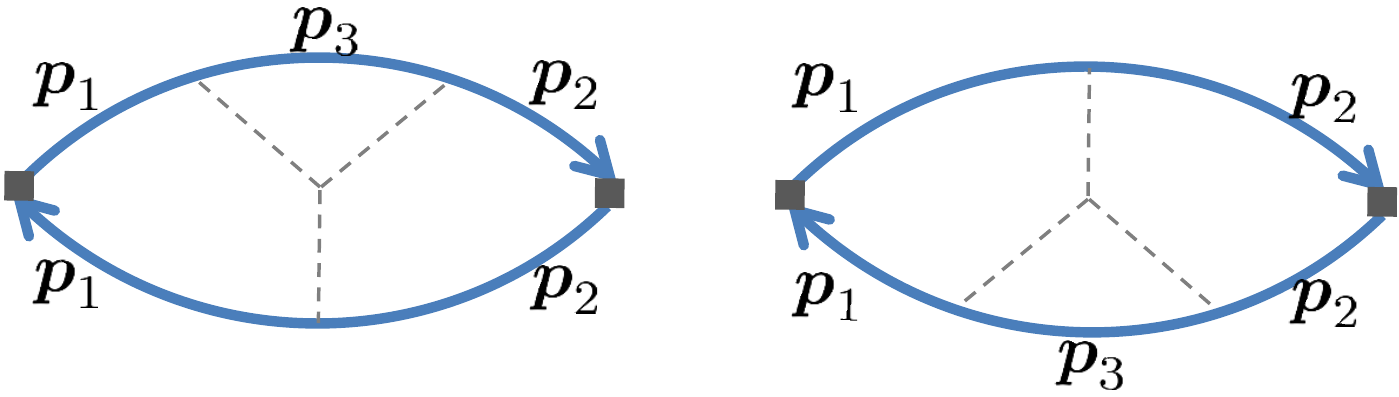}
		\caption{The Feynman diagrams for the response function $\Pi^{\mathbf{I}}_{\alpha\beta}$ due to the third order impurity scatterings. The thick solid lines are the Green's function in the spin basis under the Born approximation.}\label{fig:third_order}
\end{figure}

Though we mainly focus on the Gaussian disorder in this work, we also have a brief discussion of the skew scattering contribution due to third order impurity scatterings because this contribution is inversely proportional to the impurity density and may become significant when the  impurity density is very dilute. This contribution requires a third order correlation of the impurity potential and the corresponding Feynman diagrams are shown in Fig.\ref{fig:third_order}. We computed this contribution to the AHE in tilted Weyl metals from the Kubo-Streda formula, as shown in Appendix D, and found that the result obtained from the Kubo-Streda formula for this contribution is also the same as the result obtained from the SBE under the relaxation time approximation in the leading order of $u/v$.

There has also been an awareness   for a long time that the fourth-order crossed diagrams may give significant contribution to the AHE~\cite{Sinitsyn2007, Sinitsyn2006}. However, due to the intricacy of the calculation of these diagrams, only recently Ado {\it et al.} computed the contributions of these diagrams  for 2D Rashba ferromagnets and found that they become important for near impurities with distance comparable to the Fermi wavelength of the electrons~\cite{Ado2016}. We will present the study of these diagrams for tilted Weyl metals in a different paper.

\section{IV. Summary}

To sum up, we studied the anomalous Hall effect in disordered type-I Weyl metals with finite tilting in the Kubo-Streda formalism in the spin basis. 
We developed an efficient and transparent scheme in this basis to separate the Hall current from the three different mechanisms: intrinsic, side jump and skew scattering. 
This scheme is applicable for general relativistic systems, such as Weyl and Dirac models, both isotropic and anisotropic.
We compared the anomalous Hall current for tilted Weyl metals obtained in this way with the results from the SBE approach and found that in the leading order of the tilting velocity, the results from the two approaches agree well with each other.
Our scheme is especially important for studying the AHE in anisotropic systems since both the SBE approach and the Kubo-Streda formula in the chiral basis encounter difficulty in studying the AHE for such systems.

\section{Acknowledgements}
We thank Rui Wang for helpful discussions. This work is supported by the National Natural Science Foundation of China under Grant No. 11974166.

\begin{widetext}

\section{Appendix}

\subsection{A. Calculation of the ${\cal I}$ matrix}
In this appendix, we show the calculation of the anti-symmetric part of the polarization operator ${\cal I}$. The calculation of the symmetric part is similar. Since the final result of the dc anomalous Hall conductivity only depends on the parameters of the ${\cal I}$ matrix in the dc limit, for simplicity we only show the calculation in the dc limit in this appendix. 

 The polarization operator in the dc limit is
 \begin{equation}\label{eq:I_matrix}
{\cal I}_{\alpha\beta}(\omega\to 0, \mathbf q\to 0)=\frac{1}{2}\int\frac{d^3{\mathbf k}}{(2\pi)^3}\rm Tr[\sigma^{\alpha}G^R(\epsilon+\omega, \mathbf k+ \mathbf q)\sigma^{\beta}G^A(\epsilon,\mathbf k)].
\end{equation}

The Green's function of the tilted Weyl metals under the first Born Approximation is 
\begin{equation}
G^{R/A}(\omega \to 0, {\mathbf q \to 0})=\frac{1}{(\epsilon-\mathbf u \cdot \mathbf k -v k\pm\frac{i}{2\tau^{+}})(\epsilon-\mathbf u \cdot \mathbf k +v k\pm\frac{i}{2\tau^{-}})}[(\epsilon\pm\frac{i}{2\tau}-\mathbf u \cdot \mathbf k)\sigma^0+v\mathbf k \cdot \mathbf{\boldsymbol\sigma} \mp\frac{i}{2\tau}(\mathbf \Delta \cdot \mathbf{\boldsymbol \sigma})],
\end{equation}
where ${\mathbf \Delta}=-{\mathbf u}/v$, $\tau$ is given in the main text and $1/\tau^{\pm}=\frac{1}{\tau}(1\pm \frac{{\mathbf k}\cdot {\mathbf \Delta}}{k})$.

The numerator of the integrand of ${\cal I}_{\alpha\beta}$ is 
\begin{equation}
N=\frac{1}{2}\rm Tr\{\sigma^{\alpha}[(\epsilon+\frac{i}{2\tau}-\mathbf u \cdot \mathbf k)\sigma_0+v\mathbf k \cdot \mathbf{\boldsymbol\sigma}-\frac{i}{2\tau}(\mathbf \Delta \cdot \mathbf{\boldsymbol \sigma})]\sigma^{\beta}[(\epsilon-\frac{i}{2\tau}-\mathbf u \cdot \mathbf k)\sigma_0+v\mathbf k \cdot \mathbf{\boldsymbol \sigma}+\frac{i}{2\tau}(\mathbf \Delta \cdot \mathbf{\boldsymbol \sigma})]\}.
\end{equation}
We separate its  antisymmetric part into two parts:
\begin{equation}
(1):\ \frac{1}{2}v\ \rm Tr[(\epsilon+\frac{i}{2\tau}-\mathbf u \cdot \mathbf{\boldsymbol k})\sigma^{\alpha}\sigma^{\beta}(\mathbf k \cdot \mathbf{\boldsymbol \sigma})+(\epsilon-\frac{i}{2\tau}-\mathbf u \cdot \mathbf k)\sigma^{\alpha}(\mathbf k \cdot \mathbf{\boldsymbol \sigma})\sigma^{\beta}],
\end{equation}

\begin{equation}
(2):\ \frac{1}{2}(\frac{i}{2\tau})\ \rm Tr[(\epsilon+\frac{i}{2\tau}-\mathbf u \cdot \mathbf k)\sigma^{\alpha}\sigma^{\beta}(\mathbf \Delta \cdot \mathbf{\boldsymbol \sigma})-(\epsilon-\frac{i}{2\tau}-\mathbf u \cdot \mathbf k)\sigma^{\alpha}(\mathbf \Delta \cdot \mathbf{\boldsymbol \sigma})\sigma^{\beta}].
\end{equation}

Note that the term $\rm Tr[\sigma^\alpha(\mathbf k \cdot \mathbf{\boldsymbol \sigma})\sigma^{\beta}(\mathbf \Delta \cdot \mathbf{\boldsymbol \sigma})-\sigma^\alpha(\mathbf \Delta \cdot \mathbf{\boldsymbol \sigma})\sigma^{\beta}(\mathbf k \cdot \mathbf{\boldsymbol \sigma})]$ also contains an anti-symmetric part, but the integration over this anti-symmetric part vanishes.
Since ${\rm Tr}(\sigma^{\alpha}\sigma^{\beta}\sigma^{\gamma})=2i\epsilon^{\alpha\beta\gamma}$, parts (1) and (2) can be simplified as 
\begin{equation}\label{eq:clean_part}
(1)=\frac{1}{2}v\times \frac{i}{\tau}k^{\gamma}Tr(\sigma^{\alpha}\sigma^{\beta}\sigma^{\gamma})=-\frac{1}{\tau}\epsilon^{\alpha\beta\gamma}v k^{\gamma},
\end{equation}
\begin{equation}\label{eq:impurity_part}
(2) =(\frac{i}{2\tau})(\epsilon-\mathbf u \cdot \mathbf k)\Delta^{\gamma}  Tr(\sigma^{\alpha}\sigma^{\beta}\sigma^{\gamma})=-\frac{1}{\tau}(\epsilon-\mathbf u \cdot \mathbf k)\epsilon^{\alpha\beta\gamma} \Delta^{\gamma}.
\end{equation}

\ 

The denominator of the integrand of ${\cal I}$ is
\begin{eqnarray}
D&=&\frac{1}{(\epsilon-\mathbf u \cdot \mathbf k -v k+\frac{i}{2\tau^{+}})(\epsilon-\mathbf u \cdot \mathbf k +v k+\frac{i}{2\tau^{-}})} \frac{1}{(\epsilon-\mathbf u \cdot \mathbf k -v k-\frac{i}{2\tau^{+}})(\epsilon-\mathbf u \cdot \mathbf k +v k-\frac{i}{2\tau^{-}})}\nonumber\\
&=&\frac{1}{(\epsilon-\mathbf u \cdot \mathbf k -v k)^2+(\frac{1}{2\tau^+})^2} \frac{1}{(\epsilon-\mathbf u \cdot \mathbf k +v k)^2+(\frac{1}{2\tau^-})^2}.
\end{eqnarray}

For  $\epsilon_F \tau \gg 1$, $1/\tau^-$ is negligible,
\begin{equation}
  D\approx 2\pi \tau^+ \delta(\epsilon-\mathbf u \cdot \mathbf k -vk)\times \frac{1}{4v^2 k^2}.
\end{equation}

The total anti-symmetric part of the ${\cal I}$ matrix is then
\begin{align}\label{polorization operator}
  \mathcal{I}^a
  &=\int \frac{d^3k}{(2\pi)^3}\frac{1}{4v^2 k^2}\delta(\epsilon-\mathbf u \cdot \mathbf k -v k)\times 2\pi \tau^+(-\frac{1}{\tau})\epsilon^{\alpha\beta\gamma}[v k^{\gamma}+(\epsilon-\mathbf u \cdot \mathbf k)\Delta^{\gamma}]\nonumber\\
  &=-\frac{1}{(2\pi)^2}\epsilon^{\alpha\beta\gamma}\int d^3 k\ \frac{1}{4v k^2}\delta(\epsilon-\mathbf u \cdot \mathbf k -v k) \frac{\tau^+}{\tau}(k^{\gamma}+k\Delta^{\gamma})\nonumber\\
  &=-\frac{1}{(2\pi)^2}\epsilon^{\alpha\beta\gamma}\int d^3 k\ \frac{1}{4v k^2}\delta(\epsilon-\mathbf u \cdot \mathbf k -v k) \frac{1}{1+(\mathbf k \cdot \mathbf \Delta /k) }(k^{\gamma}+k\Delta^{\gamma})
\end{align}

To do the above integration, we rotate the $z$-axis to the direction of the vector $\mathbf u$.
Suppose $\mathbf u=u(\sin\alpha\cos\beta \cdot \hat{\mathbf x}+\sin\alpha \sin \beta\cdot\hat{\mathbf y}+\cos\alpha \cdot \hat{\mathbf z})$ in the old coordinate. The rotation to the new coordinate is 
\begin{equation}
  \left(
  \begin{array}{c}
   \hat{\mathbf x'} \\
    \hat{\mathbf y'} \\
   \hat{\mathbf z'}
  \end{array}\right)=\left(
  \begin{array}{ccc}
    \cos \alpha \cos \beta & \cos \alpha \sin \beta & -\sin \alpha \\
    -\sin \beta & \cos \beta & 0 \\
    \sin \alpha \cos \beta & \sin \alpha \sin \beta & \cos \alpha
  \end{array}\right) \
  \left( \begin{array}{c}
   \hat{\mathbf x} \\
    \hat{\mathbf y}\\
    \hat{\mathbf z}
         \end{array}\right).
\end{equation}

Now we have $\mathbf u=u\hat{\mathbf z'}$. Suppose $\mathbf k=k\ (\sin \theta \cos \phi \cdot \hat{\mathbf x'} +\sin \theta \sin \phi \cdot \hat{\mathbf y'} +\cos \theta \cdot \hat{\mathbf z'}),\ \mathbf u \cdot \mathbf k=uk\cos \theta$ in the new coordinate. In the old coordinate,
\begin{equation}
\left\{\begin{array}{ll}
k_x=&k(\cos \alpha \cos \phi \sin \theta\cos \beta- \sin \theta \sin \phi\sin\beta +\cos \theta \sin \alpha\cos\beta), \\
k_y=&k (\cos \alpha \cos \phi \sin \theta\sin \beta+ \sin \theta \sin \phi\cos\beta +\cos \theta \sin \alpha\sin\beta),\\
k_z=&k(-\sin \theta \cos \phi \sin \alpha+\cos \theta \cos \alpha).\\
                     \end{array}\right.\\
  \end{equation}                   
We then have $\int_{0}^{2\pi}d\phi \ k^{\gamma}=2\pi k\cos\theta\frac{u^{\gamma}}{u},\ \delta(\epsilon-\mathbf{ u \cdot k } -v k)=\frac{1}{u\cos\theta+v}\delta(k-\frac{\epsilon}{u\cos\theta+v})$. 

The intrinsic part of ${\cal I}^a$ is obtained by setting $\tau\to \infty$ and $\Delta=0$. We then get
\begin{align}\label{intrinsic contribution}
  \mathcal{I}_{int}^a
  & = -\lim_{\tau\rightarrow \infty}\int \frac{d^3 k}{(2\pi)^3}\frac{1}{4v^2k^2}\delta(\epsilon-\mathbf u \cdot \mathbf k -v k)\times 2\pi \frac{\tau^+}{\tau}\epsilon^{\alpha\beta\gamma}v k^{\gamma}\nonumber\\
  & = -\int \frac{d^3 k}{(2\pi)^3}\frac{1}{4v^2 k^2}\delta(\epsilon-\mathbf u \cdot \mathbf k -vk)\times 2\pi\epsilon^{\alpha\beta\gamma}v k^{\gamma}\nonumber \\
  & = -\frac{\epsilon^{\alpha\beta\gamma}}{2\pi}\frac{u^{\gamma}}{u}\int_{0}^{\infty}k^2 dk\int_{0}^{\pi}\sin\theta d\theta  k\cos\theta\frac{1}{u\cos\theta+v}\delta(k-\frac{\epsilon}{u\cos\theta+v})\frac{1}{4v k^2} \nonumber\\
  & =\frac{\epsilon_F}{4v^4\pi}\epsilon^{\alpha\beta\gamma}u^{\gamma} [\frac{v^2}{v^2-u^2}-a(u)],
 \end{align}
where $a(u)=\frac{v^3}{2u^3}\ln\frac{v+u}{v-u}-\frac{v^2}{u^2}=\frac{1}{3}+{\cal O}(u^2/v^2)$.

\

The remaining part of ${\cal I}^a$ is ${\cal I}^a_{im}$, i.e., 
\begin{eqnarray}\label{impurity contribution}
\mathcal{I}_{im}^a
   &=&\mathcal{I}^a-\mathcal{I}_{int}^a
   = -\int \frac{d^3 k}{(2\pi)^3}\frac{1}{4v^2 k^2}\delta(\epsilon-\mathbf u \cdot \mathbf k -v k)\times 2\pi \epsilon^{\alpha\beta\gamma}[(\frac{\tau^+}{\tau}-1)v k^{\gamma}+\frac{\tau^+}{\tau}(\epsilon-\mathbf u \cdot \mathbf k)\Delta^{\gamma}] \nonumber\\
   &=&-\frac{\epsilon}{8 \pi v^3}\epsilon^{\alpha\beta\gamma}\Delta_{\gamma}[1+a(u)+\frac{u^2}{v^2}a(u)].
\end{eqnarray}

We can compute the symmetric part of the ${\cal I}$ matrix similarly.  The resulting ${\cal I}^s$ matrix is shown in the main text.

\subsection{B. Connection between the ${\cal I}$ matrix in the spin basis and diagrams in the chiral basis}

In this appendix, we show the correspondence between the symmetric and anti-symmetric parts of the ${\cal I}$ matrix in the spin basis and the Feynman diagrams in the chiral basis.

The integrand $I_{\alpha\beta}$ of the ${\cal I}$ matrix in the band eigenstate basis is expanded as in Eq.(\ref{eq:I_integrand}) in the main text. We denote the elements of the GF in the eigenstate basis under the Born approximation as  $\dirac{+}{G}{+}=G^+,\dirac{-}{G}{-}=G^-,\dirac{+}{G}{-}=G^{+-},\dirac{-}{G}{+}=G^{-+}$, and the same notation for $\hat{j}_\alpha$.

The first two terms of $I_{\alpha\beta}$ in Eq.(\ref{eq:I_integrand}) are 
\begin{equation}
 I^s_{\alpha\beta} =  j^{++}_{\alpha}G^{R+}j^{++}_{\beta}G^{A+}+j^{--}_{\alpha}G^{R-}j^{--}_{\beta}G^{A-}.
\end{equation}
It is obvious that $I^s_{\alpha\beta}$ is symmetric under the exchange of $\alpha$ and $\beta$.
Since $G^{R+}G^{A+}\approx 2\pi \tau^+\delta(\epsilon-\epsilon_{+})$ and  $G^{R-}G^{A-}\approx 2\pi \tau^-\delta(\epsilon-\epsilon_{-})$, for Fermi energy $\epsilon>0$, $G^{R-}G^{A-}\sim 0$. The dominant contribution of ${\cal I}_{\alpha\beta}^s$ then only contains the first term with upper band scattering. The symmetric matrix ${\cal I}_s$ then corresponds to the diagram in Fig.\ref{fig:I_diagram}(a).

\

The next two terms in Eq.(\ref{eq:I_integrand}) are $I^{(2)}_{\alpha\beta}=j^{+-}_{\alpha}G^{R-}j^{-+}_{\beta}G^{A+}+j^{-+}_{\alpha}G^{R+}j^{+-}_{\beta}G^{A-}$. We show below that its dominant part is anti-symmetric.
Since $j^{+-}_{\alpha}=(j^{-+}_{\alpha})^*, G^{R-}G^{A+}=(G^{R+}G^{A-})^*$, one can write 
\begin{equation}
G^{R-}G^{A+}=X+iY,\ G^{R+}G^{A-}=X-iY,\ j^{+-}_{\alpha}=\rm  Re\ j^{+-}_{\alpha}+i\  Im\ j^{+-}_{\alpha}, \ j^{-+}_{\beta}=\ Re\ j^{-+}_{\beta}+i \ \rm Im\ j^{-+}_{\beta},
\end{equation}
and 
\begin{eqnarray}\label{eq:anti_symmetric}
&&\ \ \ j^{+-}_{\alpha}G^{R-}j^{-+}_{\beta}G^{A+}+j^{-+}_{\alpha}G^{R+}j^{+-}_{\beta}G^{A-} \nonumber\\
&&=(\rm Re\ j^{+-}_{\alpha}+i\ Im\ j^{+-}_{\alpha})( Re\ j^{-+}_{\beta}+i\ Im\ j^{-+}_{\beta})(X+iY)+(Re\ j^{+-}_{\alpha}-i\ Im\ j^{+-}_{\alpha})(Re\ j^{-+}_{\beta}-i\ Im\ j^{-+}_{\beta})(X-iY)\nonumber\\
&&=2 X(\rm Re\ j_{\alpha}^{+-}Re\ j^{-+}_{\beta}-Im\ j_{\alpha}^{+-}Im\ j^{-+}_{\beta} )-2 Y(Im\ j_{\alpha}^{+-}Re\ j^{-+}_{\beta}+Re\ j_{\alpha}^{+-}Im\ j^{-+}_{\beta}).
\end{eqnarray}

Since $\rm Re\ j^{-+}_{\alpha}=Re\ j^{+-}_{\alpha},Im\ j^{-+}_{\alpha}=-Im\ j^{+-}_{\alpha}$, the first term in Eq.(\ref{eq:anti_symmetric}) is symmetric and the second term is anti-symmetric.
However, since $X=Re[G^{R+}G^{A-}]$ is smaller than $G^{R+}G^{A+}$ by a factor $1/\epsilon_F\tau$, the symmetric part of this equation is negligible compared to the symmetric part in ${\cal I}^s$. We then only need to keep the anti-symmetric part $-2{\rm Im} [G^{R+}G^{A-}]({\rm Im}\ j_{\alpha}^{+-}{\rm Re}\ j^{-+}_{\beta}+{\rm Re}\ j_{\alpha}^{+-}{\rm Im}\ j^{-+}_{\beta})$ of $I^{(2)}_{\alpha\beta}$. 

In $I^{(2)}_{\alpha\beta}$, we can expand $G^{R/A}=G_0^{R/A}+G_0^{R/A} \Sigma^{R/A} G^{R/A}$. Since the self-energy $\Sigma$ introduces an extra small parameter $1/\tau$ which cannot be compensated by another $\tau$, the dominant contribution of $I^{(2)}_{\alpha\beta}$ is equal to the integration of  $j^{+-}_{\alpha}G_0^{R-}j^{-+}_{\beta}G_0^{A+}+j^{-+}_{\alpha}G_0^{R+}j^{+-}_{\beta}G_0^{A-}$.
Since this is the only nonvanishing antisymmetric part in the clean limit, it is equal to ${\cal I}^a_{int}$ in the spin basis, and corresponds to the two diagrams in the chiral basis in Fig.\ref{fig:I_diagram}c in the main text.

\

Among the remaining eight terms in $I_{\alpha\beta}$, the four terms with $G^{A-}$ are smaller in $1/\epsilon_F \tau$ and so are neglected. The remaining four terms contain $G^{+-}$ or $G^{-+}$ and $G^+$. One can expand $G^{+-}=G^{+-}_0 +  G_0^{+}\Sigma G^-\approx G^{+-}_0 + G_0^{+}\Sigma G_0^- $. It is easy to check that the replacement in the last equation does not change the dominant contribution of the integration over $I_{\alpha\beta}$. Since $G^{+-}_0=0$ in the eigenstate basis, and the dominant contribution can have only one $G^-$ line~\cite{Sinitsyn2007}, the dominant four terms with $G^{+-}$ or $G^{-+}$ correspond to the four diagrams in Fig.\ref{fig:I_diagram}(d).

The integrand of the four diagrams in Fig.\ref{fig:I_diagram}(d) can be written as 
\begin{eqnarray}
  I^{(3)}_{\alpha\beta} &&= j_{\alpha}^{++}j_{\beta}^{-+}G_0^{R+}\Sigma^R G_0^{R-}G^{A+} +j_{\beta}^{++}j_{\alpha }^{-+}G_0^{A+}\Sigma^A G_0^{A-}G^{R+}\nonumber\\
  &&+\  j_{\alpha}^{++}j_{\beta}^{+-}G^{R+}G_0^{A-}\Sigma^A G_0^{A+}+ j_{\beta}^{++}j_{\alpha}^{+-}G^{A+}G_0^{R-}\Sigma^R G_0^{R+}.
\end{eqnarray}
Since $G_0^{R+}\Sigma^R G_0^{R-}G^{A+}=(G_0^{A+}\Sigma^A G_0^{A-}G^{R+})^*, G^{R+}G_0^{A-}\Sigma^A G_0^{A+}=(G^{A+}G_0^{R-}\Sigma^R G_0^{R+})^*$, the symmetric part of $I^{(3)}_{\alpha\beta}$ is 
\begin{equation}
   I^{3s}_{\alpha\beta} =(j_{\alpha}^{++}j_{\beta}^{-+}+j_{\beta}^{++}j_{\alpha }^{-+}){\rm Re}(G_0^{R+}\Sigma^R G_0^{R-}G^{A+})+( j_{\alpha}^{++}j_{\beta}^{+-}+j_{\beta}^{++}j_{\alpha}^{+-}){\rm Re}(G^{R+}G_0^{A-}\Sigma^A G_0^{A+}),
\end{equation}
and the antisymmetric part is 
\begin{equation}
   I^{3a}_{\alpha\beta} =i(j_{\alpha}^{++}j_{\beta}^{-+}-j_{\beta}^{++}j_{\alpha }^{-+}){\rm Im}(G_0^{R+}\Sigma^R G_0^{R-}G^{A+})+i(j_{\alpha}^{++}j_{\beta}^{+-}-j_{\beta}^{++}j_{\alpha}^{+-}){\rm Im}(G^{R+}G_0^{A-}\Sigma^A G_0^{A+}).
\end{equation}

However, the symmetric part  $I^{3s}_{\alpha\beta}$ is smaller than the symmetric part $I^s_{\alpha\beta}$ by a factor $1/\epsilon_F \tau$.
So we only need to keep the antisymmetric part of $I^{(3)}_{\alpha\beta}$, which is equal to $I^{a}_{im}$ obtained in the spin basis since this antisymmetric part is nonvanishing only with impurity scattering.

\subsection{C.The intrinsic, side jump and skew scattering contributions from the SBE approach under the relaxation time approximation}

In this appendix, we redo the calculation of the intrinsic, side jump and skew scattering contributions in tilted Weyl metals from the SBE approach under the relaxation time approximation (RTA) in Ref.\cite{Fu2021}.

The Hamiltonian of the three-dimensional (3D) Weyl metal in each valley in our main text is $H_{\chi}=\chi v_F \mathbf{\boldsymbol\sigma} \cdot \mathbf k + \mathbf u_{\chi} \cdot \mathbf k$, with $\chi=\pm 1$ and ${\mathbf u}_+=-{\mathbf u}_-$. As we show in the main text, in this case the contributions to the AHE in the two valleys add up instead of cancel out. 
In the following, we only need to compute the different contributions to the AHE for a single valley and double the result at the end for two valleys.

The two eigenstates of $H_\chi$ with $\chi=+1$ are
\begin{equation}\label{eq:eigenstate}
\ket{u^+_k}=\left(\begin{array}{c}
                \cos(\theta/2) \\
                \sin(\theta/2)e^{i\phi}
              \end{array}\right),            
\ket{u^-_k}=\left(\begin{array}{c}
                \sin(\theta/2) \\
                -\cos(\theta/2)e^{i\phi}
              \end{array}\right),
\end{equation}
with $\cos\theta=k_z/k,\tan\phi=k_y/k_x$. 

\

{\it Intrinsic contribution}. The intrinsic Hall conductivity for a single valley is 
\begin{equation}\label{eq:intrinsic_conductivity}
\sigma^{int}_{xy}= e^2\int_{k<k_F}\frac{d^3k}{(2\pi)^3} \mathbf\Omega^+(k)=\frac{e^2\epsilon_F}{12\pi^2v^2}u,
\end{equation}
where the Berry curvature of the upper band is $\Omega^+(k)=-\frac{{\mathbf k}}{2k^3}$. The total intrinsic contribution from the two valleys doubles and is $\sigma^{int}_{xy}=\frac{e^2\epsilon_F}{6\pi^2v^2}u$, consistent with the value in Ref~\cite{Fu2021}.

\

{\it Side jump contribution.} There are two different mechanisms for the side jump contribution. One is directly due to the transverse coordinate shift 
\begin{equation}\label{eq:coordinate_shift}
\delta \mathbf r_{\mathbf k, \mathbf k'}=\inner{u_{k'}^+}{i\frac{\partial}{\partial \mathbf k'}u_{k}^+}-\inner{u_{k}^+}{i\frac{\partial}{\partial \mathbf k}u_{k}^+}-(\frac{\partial}{\partial \mathbf k'}+\frac{\partial}{\partial \mathbf k})\arg(\inner{u^+_{k'}}{u^+_k}).
\end{equation}

The other is due to the anomalous distribution function $g^a_k$ resulting from the  coordinate shift $\delta \mathbf r_{k,k'}$. The two contributions correspond to symmetric diagrams in the chiral basis and so should have the same value~\cite{Sinitsyn2007}. Both contributions come from the second order symmetric impurity scattering $\omega^{(2)}_{k, k'}$. In the following, we compute the two contributions  respectively from the SBE approach.

From Eqs.(\ref{eq:eigenstate}) and (\ref{eq:coordinate_shift}), we get 
\begin{equation}\label{eq:coordinate_shift_2}
\delta\mathbf r_{\mathbf k, \mathbf k'}=\frac{1}{4}\frac{k'+k}{k'^2k^2}({\mathbf k'}\times {\mathbf k})\times\frac{1}{|\inner{u^+_{k'}}{u^+_k}|^2}.
\end{equation}
For systems with isotropic Fermi surface, $k=k'$, Eq.(\ref{eq:coordinate_shift_2}) reduces to $\delta \mathbf r_{k,k'}=\frac{\mathbf\Omega^+(k)\times (\mathbf k'-\mathbf k)}{|\inner{u_{k'}^+}{u^+_k}|^2}$, where $\Omega^+(k)=-\frac{{\mathbf k}}{2k^3}$ is the Berry curvature of the upper band of the tilted Weyl metals. This is the form used in Ref.\cite{Fu2021} to compute the side jump  contribution in tilted Weyl metals. However, the anisotropy of the Fermi surface plays an important role for tilted Weyl metals and the coordinate shift we obtained in Eq.(\ref{eq:coordinate_shift_2}) after taking into account this anisotropy  gives a different side-jump velocity from the isotropic formula.

The  side-jump velocity is
\begin{equation}
    \mathbf v^{sj}_{\mathbf k}=\sum_{k'}\omega^{(2)}_{\mathbf k, \mathbf k'}\delta \mathbf r_{\mathbf k, \mathbf k'},
\end{equation}
where the second order symmetric scattering rate $\omega^{(2)}_{\mathbf k, \mathbf k'}=2\pi |V_{\mathbf k, \mathbf k'}|^2 \delta(\epsilon_k-\epsilon_{k'})$ with $|V_{\mathbf k, \mathbf k'}|^2=n_iV_0^2|\inner{u^+_{k'}}{u^+_k}|^2$. 
 The side jump velocity is then 
 \begin{equation}
\mathbf v_{\mathbf k}^{sj}=-\frac{1}{4}\frac{\mathbf k}{k^2}\times 2\pi n_iV_0^2\sum_{k'}\delta(\epsilon_k-\epsilon_{k'})\frac{k+k'}{k'^2}\mathbf k'.
   \end{equation}

For simplicity, we assume the tilting in the $z$ direction, i.e. ${\mathbf u}=(0, 0, u)$. We then get 
\begin{eqnarray}
   \mathbf v_{\mathbf k}^{sj}
  &=&-\frac{1}{4}\frac{\mathbf k\times \hat{z}}{k^2}\times 2\pi n_iV_0^2\frac{1}{(2\pi)^3}\int_{0}^{\infty}dk'\int_{0}^{\pi}\sin\theta'd\theta'\int_{0}^{2\pi}d\phi'\delta(\epsilon_k-\epsilon_{k'})(k+k')k'_z\\
   &\approx& \frac{5n_iV_0^2}{12\pi}\frac{u}{v^2}(\mathbf k\times \hat{z}). \label{eq:sj_velocity}
   \end{eqnarray}
In the last line, we have only kept the leading order of $u/v$. The above result does not depend on whether $\mathbf{u}$ is in the z direction, so $\mathbf v_k^{sj}\sim \mathbf{k}\times \mathbf{u}$.  Note that neglecting the anisotropy of the Fermi surface of tilted Weyl metals at the calculation of $\delta\mathbf r_{k,k'}$ will result in an incorrect results  $\mathbf v_k^{sj}=\frac{2n_iV_0^2 u}{3\pi v^2} (\mathbf k\times \mathbf{u})$.

We separate the non-equilibrium distribution function as $f(\epsilon_{\mathbf k})=f_0(\epsilon_{\mathbf k})+g_{\mathbf k}+g^a_\mathbf k$, where $f_0$ is the equilibrium distribution, $g_{\mathbf k}$ is the usual non-equilibrium distribution without considering the coordinate shift and $g^a_{\mathbf k}$ is the anomalous part due to coordinate shift. The non-equilibrium parts $g_{\mathbf k}$ and $g^a_{\mathbf k}$ satisfy the following equations respectively~\cite{Sinitsyn2007}
\begin{eqnarray}
e\mathbf E \cdot \mathbf v_{\mathbf k}\frac{\partial f_0}{\partial \epsilon^+_{\mathbf k}}= -\sum_{k'}\omega_{\mathbf k, \mathbf k'}(g_{\mathbf k}-g_{\mathbf k}), \label{eq:normal_dis}\\
  \sum_{k'}\omega_{\mathbf k, \mathbf k'}(g_{\mathbf k}^a-g^a_{\mathbf k}+\frac{-\partial f_0}{\partial \epsilon^+_k}e\mathbf E \cdot \delta \mathbf r_{\mathbf k, \mathbf k'})=0, \label{eq:a_dis}
\end{eqnarray}
where ${\mathbf v_{\mathbf k}}=\frac{\partial\epsilon_k^+}{\partial \mathbf k}$.

 Assuming the electric field $\mathbf E$ is in the $xy$ plane,  $g_{\mathbf k}$ is then solved by the ansatz solution~\cite{Loss2003}
  \begin{equation}\label{eq:ne_ansatz}
g_{\mathbf k}=(-\frac{\partial f_0}{\partial\epsilon_{\mathbf k}^+})e\mathbf E \cdot (A {\mathbf v_{\mathbf k}}+B{\mathbf v_{\mathbf k}}\times \hat{{\mathbf u}}), 
\end{equation}
 where $A$ and $B$ are assumed to be constant independent of the direction of $\mathbf{k}$, and $\hat{{\mathbf u}}$ is the unit vector in the direction  of $\mathbf{u}$. 
  The solution for $g_{\mathbf k}$ under this assumption is then
\begin{equation}\label{eq:ne_dis}
g_{\mathbf k}=(-\frac{\partial f_0}{\partial\epsilon_{\mathbf k}^+})e\mathbf E \cdot ({\mathbf v_{\mathbf k}}\tau^{tr}+({\mathbf v_{\mathbf k}}\times \hat{{\mathbf u}})\tau^{tr}\tau_{\perp}^{tr}/\tau^\perp), 
\end{equation}
 where
\begin{eqnarray}
1/\tau^{tr}&=&\sum_{k'}\omega_{\mathbf k, \mathbf k'}(1-\frac{\mathbf v_{k}\cdot \mathbf v_{k'}}{|\mathbf v_{k}|^2}), \label{eq:ani_tau_tr}\\
1/\tau^\bot&=&\sum_{k'}\omega_{\mathbf k, \mathbf k'}\ \frac{\mathbf v_{k'}\cdot(\mathbf v_{k}\times \hat{{\mathbf u}})}{|\mathbf v_{k}\times \hat{{\mathbf u}}|^2}, \\
1/\tau_{\perp}^{tr}&=&\sum_{k'}\omega_{\mathbf k, \mathbf k'}(1-\frac{(\mathbf v_{k'}\times \hat{{\mathbf u}})\cdot(\mathbf v_{k}\times \hat{{\mathbf u}})}{|\mathbf v_{k}\times \hat{{\mathbf u}}|^2}). \label{eq:tau_bot_tr}
\end{eqnarray}

It is easy to check that for tilted Weyl metals, $\tau^{tr}, \tau^\bot$ and $\tau_{\perp}^{tr}$ depend on the direction of $\mathbf{k}$, contradictory with the ansatz. However, 
in the leading order of $u/v$, they reduce to the constants in untilted Weyl metals, and $1/\tau^{tr}\approx \sum_{\mathbf k'}\omega_{\mathbf k, \mathbf k'}[1-\cos(\mathbf{k}\cdot \mathbf{k'})]$, 
$1/\tau_{\perp}^{tr}=\int\frac{d^3k'}{(2\pi)^3}\omega_{\mathbf k, \mathbf k'}(1-\left|\frac{v^{\perp}_{\mathbf k'}}{v^{\perp}_{\mathbf k}}\right|\cos(\phi-\phi'))$, and 
 $1/\tau^\bot=\int\frac{d^3k'}{(2\pi)^3}\omega_{\mathbf k, \mathbf k'}\left|\frac{v^{\perp}_{\mathbf k'}}{v^{\perp}_{\mathbf k}}\right|\sin(\phi-\phi')$, where $v^\perp_{\mathbf k}=\mathbf v_{\mathbf k}\times \hat{{\mathbf u}}$ is a vector perpendicular to both  $\mathbf v_{\mathbf k}$ and $\hat{{\mathbf u}}$.

Assuming $\mathbf{E}$ is in the $y$ direction, the anomalous Hall conductivity in a single valley due to the coordinate shift is~\cite{Sinitsyn2007} 
\begin{equation}
    \sigma^{\rm sj, I}_{xy}=e\int\frac{d^3k}{(2\pi)^3}(g_{\mathbf k}/E_y)v_{{\mathbf k}, x}^{sj},
\end{equation}
where $v_{{\mathbf k}, x}^{sj}$ is the $x$ component of the side jump velocity $\mathbf{v}_{\mathbf k}^{sj}$, and  $g_{\mathbf k}$ is solved in Eq.(\ref{eq:ne_dis}).

For the symmetric second order impurity scattering $\omega^{(2)}_{k,k'}$, $1/\tau^{tr}\approx 1/\tau^{tr}_{\bot}\approx \frac{\epsilon_F^2n_iV_0^2}{3\pi v^3}$ in the leading order of $u/v$, and $1/\tau^\bot=0$, so the second term in Eq.(\ref{eq:ne_dis}) has zero contribution to the side jump contribution. 
From Eqs.(\ref{eq:sj_velocity}) and Eq.(\ref{eq:ne_dis}), we get the side jump contribution in the leading order of $u/v$ due to transverse coordinate shift for a single valley under the RTA as
\begin{equation}
\sigma^{\rm sj, I}_{xy}=\frac{5e^2\epsilon_F}{24\pi^2v^2}u_z.
\end{equation}
One can calculate $\sigma^{\rm sj, I}_{yz}$ in the same way and get the side jump current 
$j_H^{\rm sj, I}=\frac{5 e^2\epsilon_F}{24\pi^2v^2} \mathbf{E}\times\mathbf{u}$.

The second part of the side jump contribution  comes from the anomalous distribution function  due to coordinate shift. The anomalous distribution function solved from Eq.(\ref{eq:a_dis}) under the RTA is
\begin{equation}
g_{\mathbf k}^a=\frac{5n_iV_0^2u}{12\pi v^2} \frac{\partial f_0}{\partial \epsilon^+_{\mathbf k}}[e  \mathbf E\cdot(\mathbf k\times \mathbf{u})]\tau^{tr}_{\bot}.
\end{equation}
Here $1/\tau^{tr}_{\bot}$ is defined in Eq.(\ref{eq:tau_bot_tr}) in the leading order of $u/v$.

The  Hall conductivity in the leading order of $u/v$ due to the anomalous distribution for a single valley under the RTA is then 
\begin{equation}
 \sigma_{xy}^{\rm sj, II}=e\int \frac{d^3k}{(2\pi)^3}(g_k^a/E_y)v_{k, x}=\frac{5e^2\epsilon_F }{24\pi^2v^2}u_z,
\end{equation}
which is equal to the contribution due to coordinate shift. The Hall current due to this part is also $j_H^{\rm sj, II}=\frac{5 e^2\epsilon_F}{24\pi^2v^2} \mathbf{E}\times\mathbf{u}$.

It was pointed out in Ref.\cite{Sinitsyn2007} that the product of $g^a_{\mathbf k}$ and $v^{sj}_{\mathbf k}$ may give  a finite contribution to the AHE in anisotropic systems. 
However, it is easy to check that under the RTA, this contribution to the Hall current is zero in the leading order of $u/v$.

The total side jump contribution under the RTA for two valleys is then 
\begin{equation}
j_H^{\rm sj}=\frac{5 e^2\epsilon_F}{6\pi^2v^2} \mathbf{E}\times\mathbf{u}.
\end{equation}

\

{\it Skew scattering contribution.}
The skew scattering contribution is due to the asymmetric fourth-order impurity scatterings. Still assuming the electric field $\mathbf{E}$ in the $y$ direction, the Hall conductivity from the skew scattering contribution for a single valley is~\cite{Sinitsyn2007} 
\begin{equation}
    \sigma^{sk}_{xy}=e\sum_{\mathbf k}(g_{\mathbf k}/E_y)v_{{\mathbf k},x},
\end{equation}
where the solution of $g_{\mathbf k}$ under RTA  is still Eq.(\ref{eq:ne_dis}) but with the contribution of fourth-order scattering $\omega_{\mathbf k, \mathbf k'}^{(4)}=-\frac{2\epsilon_F}{3v^4}(n_iV_0^2)^2 u\sin\theta\sin\theta'\sin(\phi-\phi')\delta(\epsilon_{\mathbf k}-\epsilon_{\mathbf k'})$ taken into account. The contribution of $\omega_{\mathbf k, \mathbf k'}^{(4)}$ to $1/\tau^{tr}$ and $1/\tau^{tr}_{\bot}$ is zero, but $1/\tau^{\bot}=-\frac{1}{9}\frac{\epsilon^3_F u}{\pi^2 v^7}(n_iV_0^2)^2$ is non-zero for $\omega_{\mathbf k, \mathbf k'}=\omega_{\mathbf k, \mathbf k'}^{(4)}$. The skew scattering contribution under RTA is then 
\begin{equation}
\sigma^{sk}_{xy}=-\int \frac{d^3k}{(2\pi)^3}\left[(\frac{\partial f_0}{\partial\epsilon_{\mathbf k}^+})e\mathbf E \cdot ({\mathbf v_{\mathbf k}}\times \hat{{\mathbf u}})\tau^{tr}\tau_{\perp}^{tr}/\tau^\perp\right] v_{\mathbf k, x}/E_y,
\end{equation}
where the contribution to $1/\tau^{tr}$ and $1/\tau^{tr}_\bot$ still comes from the second order impurity scatterings.

Putting all things together, we get the leading order total skew scattering contribution under RTA for two valleys as 
\begin{equation}
j_H^{\rm sk}=\frac{ e^2\epsilon_F}{3\pi^2v^2} \mathbf{E}\times\mathbf{u}.
\end{equation}
which is the same as the value obtained in Ref.~\cite{Fu2021}.

The sign of the intrinsic contribution we obtained is opposite to that of Ref.~\cite{Fu2021} since we use a different convention in Eq.(\ref{eq:intrinsic_conductivity}). 
The signs of the three contributions we obtained are then the same as the signs we obtained from the  Kubo-Streda formula.

\subsection{D.The skew-scattering contribution due to third order impurity scatterings}

We compute the anomalous Hall effect due to third order skew scatterings of impurities by the Kubo-Streda formula in this appendix. For simplicity, we only keep the leading order of $u/v$ in this section.

The diagrams corresponding to these processes are shown in Fig.\ref{fig:third_order}. We first consider the response function without the vertex correction at the two ends. The corresponding response function for the two diagrams is 
\begin{eqnarray}\label{eq:third_order_response}
  \Pi_{\alpha\beta}^{\rm bare}=&& e^2v^2n_iV_1^3 \frac{i\omega}{2\pi}\sum_{\mathbf k_1}\sum_{\mathbf k_2}\sum_{\mathbf k_3}
\{ {\rm Tr}[G^A(\epsilon, \mathbf k_1)\sigma_{\alpha}G^R(\epsilon, \mathbf k_1)G^R(\epsilon, \mathbf k_3)G^R(\epsilon, \mathbf k_2)\sigma_{\beta}G^A(\epsilon, \mathbf k_2)] \nonumber\\
  &&\ \ \ \ \  \ \ \ \ \ \ \ \ \ \ \  \ \ \  \ \ \ \ \ \ \ \ \ \ \ +{\rm Tr}[G^A(\epsilon, \mathbf k_1)\sigma_{\alpha}G^R(\epsilon, \mathbf k_1)G^R(\epsilon, \mathbf k_2)\sigma_{\beta}G^A(\epsilon, \mathbf k_2)G^A(\epsilon, \mathbf k_3)]\},
\end{eqnarray}
where the third order distribution $\langle V_i^3 \rangle =V^3_1$ for the impurity potential. The energy in the Green's function is bounded to the Fermi energy, i.e.,  $\epsilon=\epsilon_F$.

The integrand of Eq.(\ref{eq:third_order_response}) can be related to the integrand of the polarization matrix as 
\begin{equation}
G^A(\epsilon, \mathbf k_1)\sigma_{\alpha}G^R(\epsilon, \mathbf k_1)=I_{\alpha\mu}(\epsilon, \mathbf k_1)\sigma_{\mu},
\ G^R(\epsilon, \mathbf k_2)\sigma_{\beta}G^A(\epsilon, \mathbf k_2)=\sigma_{\nu} I_{\nu\beta}(\epsilon, \mathbf k_2)
\end{equation}
where $I_{\alpha\mu}(\epsilon, \mathbf k)=\frac{1}{2}\rm Tr[\sigma_{\alpha}G^R(\epsilon, \mathbf k)\sigma_{\mu}G^A(\epsilon, \mathbf k)]$.

Defining
\begin{equation}
f_{\mu\nu}(\epsilon, \mathbf k_3)={\rm Tr}[\sigma_{\mu}G^R(\epsilon, \mathbf k_3)\sigma_{\nu}]+{\rm Tr}[\sigma_{\mu}\sigma_{\nu}G^A(\epsilon, \mathbf k_3)],
\end{equation}
the response function in Eq.(\ref{eq:third_order_response}) then becomes
\begin{align}\label{11}
  \Pi_{\alpha\beta}^{\rm bare}=e^2v^2n_iV_1^3  \frac{i\omega}{2\pi}\sum_{\mathbf k_1}\sum_{\mathbf k_2}\sum_{\mathbf k_3}I_{\alpha\mu}(\mathbf k_1)f_{\mu\nu}(\mathbf k_3)I_{\nu\beta}(\mathbf k_2)
  =\frac{i\omega}{2\pi} e^2v^2n_iV_1^3  {\cal I}_{\alpha\mu}{\cal F}_{\mu\nu}{\cal I}_{\nu\beta}
\end{align}
where ${\cal I}$ is the polarization matrix defined in the main text and ${\cal F}_{\mu\nu}=\sum_{\mathbf k_3}f_{\mu\nu}(\epsilon, \mathbf k_3)$.

The AHE corresponds to the anti-symmetric part of $\Pi_{\alpha\beta}$. Since the anti-symmetric part of ${\cal I}$ is smaller than its symmetric part by $1/\epsilon_F \tau$, yet the symmetric part and anti-symmetric part of ${\cal F}$ have the same order of magnitude in terms of $1/\epsilon_F \tau$, we only need to keep the symmetric part of ${\cal I}$ and the anti-symmetric part of ${\cal F}$.

The leading order anti-symmetric part of $f_{\mu\nu}(\epsilon, \mathbf k_3)$ is 
\begin{eqnarray}
 f_{\mu\nu}^a (\epsilon, \mathbf k_3)&&=2iv\epsilon_{\mu\nu \gamma}k_{3\gamma}[\frac{1}{(\epsilon-\mathbf u \cdot \mathbf k_3 -v k_3-\frac{i}{2\tau})(\epsilon-\mathbf u \cdot \mathbf k_3 +v k_3-\frac{i}{2\tau})}
  -{\rm c.c}] \nonumber\\
&&  \approx -4\pi v\epsilon_{\mu\nu \gamma}k_{3\gamma}\frac{1}{\epsilon-\mathbf u \cdot \mathbf k_3 +v k_3} \delta(\epsilon-\mathbf u \cdot \mathbf k_3 -v k_3).
\end{eqnarray}

We assume $\mathbf{u}$ in the $z$ direction, i.e., $\mathbf{u}=(0, 0, u)$ for simplicity in this appendix. The integration over $f_{\mu\nu}^a(\epsilon, \mathbf k_3)$ is then nonzero only for $\gamma=z$ in the above expression. We get 
\begin{eqnarray}
 {\cal F}_{12}= \sum_{\mathbf{k}_3}f_{12}^a(\epsilon, \mathbf k_3)
  &&=-\frac{4\pi v}{(2\pi)^3}\int_{0}^{\infty}k_3^2 dk_3\int_{0}^{\pi}\sin\theta_3 d\theta_3\int_{0}^{2\pi}d\phi_3 \delta(\epsilon-\mathbf u \cdot \mathbf k_3 -vk_3)\frac{k_{3z}}{\epsilon-\mathbf u \cdot \mathbf k_3 +vk_3} \nonumber\\
 && \approx \frac{\epsilon_F^2u}{\pi v^4}.
\end{eqnarray}

For $\mathbf{E}$ in the $y$ direction, the leading order response function corresponding to the AHE for the third order impurity scatterings is then 
\begin{equation}
\Pi^{\rm bare}_{xy}=\frac{i\omega}{2\pi} e^2v^2n_iV_1^3{\cal I}_{11} {\cal F}_{12} {\cal I}_{22}\approx \frac{i\omega}{2\pi}\frac{e^2\epsilon_F^6\tau^2u}{36\pi^3v^8}n_iV_1^3.
\end{equation}

Adding the vertex correction from the ladder diagram at the two ends of the diagrams in Fig.\ref{fig:third_order}, we get the full response function at the leading order of $u/v$ as
\begin{equation}
\Pi_{xy}^{(3)}=\frac{9}{4}\Pi^{\rm bare}_{xy}= \frac{9}{4} \frac{i\omega}{2\pi}e^2v^2n_iV_1^3{\cal I}_{11} {\cal F}_{12} {\cal I}_{22}=\frac{i\omega}{2\pi}\frac{e^2\epsilon_F^6\tau^2u}{16\pi^3v^8}n_iV_1^3.
\end{equation}

The corresponding Hall conductivity for two valleys is 
\begin{equation}\label{eq:quantum}
\sigma_{xy}^{(3)}=2\frac{\Pi^{(3)}_{xy}}{i\omega}=\frac{1}{4}\frac{e^2\epsilon_F^2 u}{\pi^2 v^2}\frac{V_1^3}{n_iV_0^4},
\end{equation}
where we have used $n_iV_0^2=\frac{2\pi v^3}{\epsilon_F^2\tau}$.

\

As a comparison, we next compute this contribution by the SBE approach under the RTA.

The Hall  conductivity from the SBE approach is 
\begin{equation}
\sigma^{(3)}_{xy}=e\sum_{\mathbf k}(g_{\mathbf k}/E_y)(\mathbf v_{\mathbf k})_x,
\end{equation}
where $g_{\mathbf k}$ is solved in Appendix  C but with $\omega_{\mathbf k, \mathbf k'}$ now includes $\omega_{\mathbf k, \mathbf k'}^{(3)}=-\frac{\epsilon^2 u}{2v^4}n_iV_1^3\sin\theta'\sin\theta\sin(\phi-\phi')\delta(\epsilon_{\mathbf k}-\epsilon_{\mathbf k'})$.

For $\sigma^{(3)}_{xy}$, only 
$g_{\mathbf k}=-\frac{\p f_0}{\p \epsilon_{\mathbf k}}e\mathbf E\cdot (\mathbf v_k\times \hat{z})(\tau^{tr})^2/\tau^{\perp}$ has non-zero contribution for $\mathbf{E}$ in the $y$ direction.

For $\omega^{(3)}_{\mathbf k, \mathbf k'}$, 
\begin{equation}
\frac{1}{\tau^{\perp}}=\sum_{\mathbf k'}\omega_{\mathbf k, \mathbf k'}^{(3)}\frac{\sin\theta'}{\sin\theta}\sin(\phi-\phi')=-\frac{\epsilon^2 u}{2v^4}n_iV_1^3\times \frac{\epsilon^2}{6\pi^2v^3}+o(u^2), 
\end{equation}

The leading order contribution to $1/\tau^{tr}$ still comes from the second order scattering and $1/\tau^{tr} =\frac{\epsilon^2n_iV_0^2}{3\pi v^3}$. We then get the Hall conductivity for two valleys due to the third order scattering as 
\begin{equation}
\sigma_{xy}^{(3)}=\frac{1}{4}\frac{e^2\epsilon_F^2u}{\pi^2 v^2}\frac{V_1^3}{n_iV_0^4}.
\end{equation}

This result is consistent with Eq.(\ref{eq:quantum}) obtained from the Kubo-Streda formula as well as the result in Ref.\cite{Fu2021}.

\end{widetext}


\begin{thebibliography}{99}
\bibitem{Footnote0}Among the eight diagrams for side jump contribution, four correspond to the contribution directly
 from the transverse coordinate shift and the other four correspond to the contribution of the anomalous distribution due to the coordinate shift. The two types have the same contributions to AHE and we call both of them the side jump contribution.
\end{thebibliography}
\end{document}